\pdfoutput=1
\documentclass[acmsmall,screen]{acmart}

\AtBeginDocument{%
  }

\startPage{1}

\setcopyright{none}
\renewcommand\footnotetextcopyrightpermission[1]{} 
\makeatletter
\renewcommand\@formatdoi[1]{\ignorespaces}
\makeatother

\def\firstfoot{\def\@firstfoot{}}
\def\firsthead{\def\@firsthead{}}

\usepackage{booktabs}   
\usepackage{subcaption} 
\usepackage{}
\usepackage{listings}
\usepackage{xspace}
\usepackage[utf8]{inputenc}
\usepackage[T1]{fontenc}
\usepackage{xcolor}
\usepackage[normalem]{ulem}
\usepackage{footnote}
\usepackage{tablefootnote}
\usepackage{pdfpages}
\usepackage{multicol}
\usepackage{multirow}
\usepackage{enumitem}
\usepackage{lstautogobble}
\usepackage{tikz}
\usetikzlibrary{calc}
\usepackage[shortcuts]{extdash}
\usepackage{amsmath}
\usepackage{newtxmath}
\usepackage{adjustbox}
\usepackage{syntax} 
\usepackage{bold-extra}
\usepackage{ifmtarg}

\usepackage{alltt}
\usepackage{algorithmicx}
\usepackage{algorithm} 
\usepackage[noend]{algpseudocode}
\usepackage{alltt}

\input{algomacros}
\usepackage{hyperref}

\usepackage{xspace}
\usepackage{tikz}
\usetikzlibrary{arrows.meta}
\usepackage{adjustbox}
\usepackage{alltt}
\usepackage{fancyvrb}
\usepackage{mathtools}
\usepackage{bold-extra}
\usepackage{lipsum}
\usepackage{lineno}
\usepackage{proof}
\usepackage{mathpartir}

\usepackage{highlighter}
\usepackage[framemethod=TikZ]{mdframed}

\newcommand{\newtextsc}[1]{\textsc{\small #1}}
\newcommand{\sysname}{\textsc{StaticFixer}\xspace}
\newcommand{\codeql}{\newtextsc{CodeQL}\xspace}
\newcommand{\othersysname}[1]{\newtextsc{#1}\xspace}
\newcommand{\sa}{\newtextsc{SA}\xspace}

\newcommand{\astree}{\newtextsc{AST}\xspace}

\newcommand{\pdg}{\newtextsc{AST}\xspace}

\newcommand{\taintprop}{information flow\xspace}

\newcommand{\taintpropedge}{datalow-edge\xspace}

\newcommand{\safeprog}{witnessing-safe program\xspace}
\newcommand{\safeprogs}{witnessing-safe programs\xspace}
\newcommand{\Safeprogs}{Witnessing-safe programs\xspace}
\newcommand{\sawitnessfull}{static-analysis witnessing\xspace}
\newcommand{\sawitness}{\sa-witnessing\xspace}

\newcommand{\js}{\newtextsc{JavaScript}\xspace}
\newcommand{\cpp}{\newtextsc{C++}\xspace}

\newcommand{\unvalidatedmembership}{\newtextsc{UDC-MembershipCheck}\xspace}
\newcommand{\unvalidatedtype}{\newtextsc{UDC-TypeCheck}\xspace}
\newcommand{\unvalidatedcall}{\newtextsc{UDC}\xspace}

\newcommand{\xss}{\newtextsc{XSS}\xspace}

\newcommand{\rmGuard}{\newtextsc{RemoveGuard}\xspace}
\newcommand{\rmSan}{\newtextsc{RemoveSanitizer}\xspace}

\newcommand{\dsl}{\newtextsc{DSL}\xspace}
\newcommand{\strategy}{\textit{S}\xspace}
\newcommand{\strategyone}{\textit{S}$_1$\xspace}
\newcommand{\strategytwo}{\textit{S}$_2$\xspace}

\newcommand{\edit}{\ensuremath{E}\xspace}
\newcommand{\insertsc}{\newtextsc{Insert}\xspace}
\newcommand{\replace}{\newtextsc{Replace}\xspace}

\newcommand{\editloc}{\textit{L}\ensuremath{_E}\xspace}
\newcommand{\editindex}{\textit{I}\ensuremath{_E}\xspace}

\newcommand{\semloc}{\textit{L}\ensuremath{_S}\xspace}
\newcommand{\refloc}{\textit{L}\ensuremath{_R}\xspace}
\newcommand{\reflocone}{\textit{L}\ensuremath{_{r1}}\xspace}
\newcommand{\refloctwo}{\textit{L}\ensuremath{_{r2}}\xspace}
\newcommand{\reflocthree}{\textit{L}\ensuremath{_{r3}}\xspace}

\newcommand{\clause}{\textit{C}\xspace}

\newcommand{\prog}{\ensuremath{P}\xspace}
\newcommand{\concedit}{\ensuremath{E}\xspace}
\newcommand{\conceditloc}{\ensuremath{\overline{\mathit{EL}}}\xspace}
\newcommand{\conceditpath}{\ensuremath{\overline{\mathit{EP}}}\xspace}

\newcommand{\concinsertcode}{\ensuremath{\mathcal{C}}\xspace}

\newcommand{\kleeneedge}{\newtextsc{KleeneEdge}\xspace}
\newcommand{\kleeneedges}{\newtextsc{KleeneEdges}\xspace}

\newcommand{\editprog}{editprog\xspace}

\newcommand{\witness}{\textit{W}\xspace}
\newcommand{\witnesspar}{\textit{W}\ensuremath{_{par}}\xspace}
\newcommand{\witnessparpar}{\textit{W}\ensuremath{_{grandpar}}\xspace}

\newcommand{\gettraversal}{\newtextsc{GetTraversal}\xspace}
\newcommand{\getclauses}{\newtextsc{GetClauses}\xspace}
\newcommand{\getindex}{\newtextsc{GetIndex}\xspace}

\newcommand{\node}{\newtextsc{Node}\xspace}
\newcommand{\bool}{\newtextsc{Bool}\xspace}
\newcommand{\integer}{\newtextsc{Integer}\xspace}
\newcommand{\eastree}{\newtextsc{E-AST}\xspace}

\newcommand{\edgetype}{\I{ET}\xspace}

\newcommand{\blockstmt}{\newtextsc{BlockStmt}\xspace}
\newcommand{\ifstmt}{\newtextsc{IfStmt}\xspace}
\newcommand{\callexpr}{\newtextsc{CallExpr}\xspace}
\newcommand{\varexpr}{\newtextsc{VarExpr}\xspace}
\newcommand{\indexexpr}{\newtextsc{IndexExpr}\xspace}
\newcommand{\dotexpr}{\newtextsc{DotExpr}\xspace}

\newcommand{\labelexpr}{\newtextsc{Label}\xspace}
\newcommand{\binaryexpr}{\newtextsc{BinaryExpr}\xspace}
\newcommand{\binaryop}{\newtextsc{BinaryOp}\xspace}
\newcommand{\expr}{\newtextsc{Expr}\xspace}
\newcommand{\assignexpr}{\newtextsc{AssignExpr}\xspace}
\newcommand{\declexpr}{\newtextsc{DeclExpr}\xspace}
\newcommand{\vardecl}{\newtextsc{VarDecl}\xspace}
\newcommand{\declarator}{\newtextsc{Declarator}\xspace}

\newcommand{\unsure}[1]{{\color{black}#1}}

\newcommand{\benchsmall}{\textbf{\newtextsc{PairedPrograms}}\xspace}
\newcommand{\benchwild}{\textbf{\newtextsc{CodeInTheWild}}\xspace}
\newcommand{\codetfive}{\newtextsc{CodeT5}}
\newcommand{\codetfivejs}{\newtextsc{CodeT5Js}}
\newcommand{\codex}{\newtextsc{Codex}\xspace}
\newcommand{\out}[1]{{}}

\newcounter{inlineenum}
\renewcommand{\theinlineenum}{\alph{inlineenum}}

\usepackage{listings}
\usepackage{color}
\definecolor{lightgray}{rgb}{.9,.9,.9}
\definecolor{darkgray}{rgb}{.4,.4,.4}
\definecolor{purple}{rgb}{0.65, 0.12, 0.82}
\definecolor{backcolour}{rgb}{0.95,0.95,0.95}
\definecolor{codegreen}{rgb}{0,0.6,0}
\definecolor{codepurple}{RGB}{138,43,226}

\lstdefinelanguage{JavaScript}{
  keywords={break, case, catch, continue, debugger, default, delete, do, else, false, finally, function, in, instanceof, new, null, return, switch, this, throw, true, try, typeof, var, void, while, with, isSource, isSink, isSanitizer, isGuard, isWitness, succ, hasFlow},
  morecomment=[l]{//},
  morecomment=[s]{/*}{*/},
  morestring=[b]',
  morestring=[b]",
  ndkeywords={class, export, boolean, throw, implements, import, this},
  keywordstyle=\color{darkgray}\bfseries,
  ndkeywordstyle=\color{darkgray}\bfseries,
  identifierstyle=\color{black},
  commentstyle=\color{codepurple}\ttfamily,
  stringstyle=\color{codepurple}\ttfamily,
  sensitive=true,
  morekeywords={isSource, isSink, isSanitizer, isGuard, isWitness, hasOwnProperty, succ, hasFlow, and, or, none, parse, run, checkSafe, executeCritical, sanitize, loadActions, serveRequest, requestListener, escape, input, ApplyTraversal, GetKleeneStar, GetClause, GetOffsetIndex, GetEdge, GetConstant, ConstantAST, ReferenceAST, Insert, Replace, has}
}

\newcommand{\I}[1]{\textit{#1}}
\newcommand{\T}[1]{\texttt{\small #1}}

\makeatletter
\newcommand{\isempty}[1]{%
  \@ifmtarg{#1}{}{\ }}
\makeatother

\newcommand{\DRulenew}[3]{\textbf{#1}\isempty{#1}\textsc{#2}\ \textit{#3}}
\makeatletter
\newcommand{\DMethod}[1]{\texttt{\small #1(}\checknextarga}
\newcommand{\DMethodDSL}[1]{\texttt{#1(}\checknextarga}
\newcommand{\checknextarga}{\@ifnextchar\bgroup{\gobblenextarga}{\texttt{)}}}
\newcommand{\checknextargb}{\@ifnextchar\bgroup{\gobblenextargb}{\texttt{)}}}
\newcommand{\gobblenextarga}[1]{#1\checknextargb}
\newcommand{\gobblenextargb}[1]{, #1\checknextargb}
\makeatother

\definecolor{ForestGreen}{RGB}{34,139,34}
\definecolor{newForestGreen}{RGB}{34,139,34}
\definecolor{newred}{RGB}{210, 43, 43}

\begin{document}

\title{\sysname : From Static Analysis to Static Repair}



\author{Naman Jain}
\author{Shubham Gandhi}
\author{Atharv Sonwane}
\author{Aditya Kanade}
\author{Nagarajan Natarajan}
\author{Suresh Parthasarathy}
\author{Sriram Rajamani}
\author{Rahul Sharma}
\affiliation{
\institution{Microsoft Research India}
\country{India}}

\begin{abstract}
Static analysis tools are traditionally used to detect and flag programs that violate properties. We show that static analysis tools
can also be used to perturb programs that satisfy a property to construct variants that violate the property. Using this insight we can construct paired data sets of unsafe-safe program pairs, and learn strategies to automatically repair property violations. 
We present a system called \sysname, which automatically repairs information flow vulnerabilities using this approach.
Since information flow properties are non-local (both to check and repair), \sysname also introduces a novel domain specific language (DSL) and strategy learning algorithms for synthesizing non-local repairs.
We use \sysname to synthesize strategies for repairing two types of information flow vulnerabilities, unvalidated dynamic calls and cross-site scripting, and show that \sysname successfully repairs several hundred vulnerabilities from open source {\sc JavaScript} repositories, outperforming neural baselines built using {\sc CodeT5} and {\sc Codex}. Our datasets can be downloaded from \url{http://aka.ms/StaticFixer}.

\end{abstract}


\settopmatter{printacmref=false}
\setcopyright{none}
\settopmatter{printfolios=true}
\maketitle

\section{Introduction}
\label{sec:intro}





Static analysis (\sa for brevity) takes a program $P$ and a property $\varphi$ as inputs, and checks if the program satisfies the property. 
If the program violates the property, \sa\ outputs an error report, which the developer uses to fix the violation.
In this paper, we present a system called \sysname\, which automates the process of fixing these violations.

Given a \sa\ tool, a property $\varphi$, and a large corpus of programs $\mathcal{P}$ with a sufficient variety of programs that satisfy $\varphi$, our system \sysname\ automatically learns to fix static analysis violations in a data-driven manner.
\sysname\ consists of two stages:
\begin{enumerate}
\item
Data collection: This stage uses the corpus $\mathcal{P}$ to construct a set of "paired" programs $\{ (P_1,P'_1), (P_2,P'_2), \ldots, (P_n,P'_n)\}$, such that, for each pair 
$(P_i,P'_i)$, the first program $P_i$ violates the property $\varphi$, the second program $P'_i$ contains the fix for the violation. Furthermore, except for the fix for the violation of $\varphi$, we have that $P_i$ and $P'_i$ are identical.
\item
Strategy learning: This stage uses the paired set of programs above as a training set to automatically learn a repair strategy \strategy for fixing the violation. Specifically, for any program, $P$ that violates $\varphi$ and roughly resembles one or more programs in $\mathcal{P}$, the goal is for $\strategy(P)$ to fix the violation of $\varphi$.
\end{enumerate}

We consider the class of information flow
safety properties. Violating these properties can result in the system becoming vulnerable to information flow attacks, allowing malicious user input to flow from an untrusted {\em source} (e.g., a server request, socket message, file upload) to a sensitive {\em sink} (e.g., dynamic function execution,
shell command, database query).
SQL Injection~\cite{sqlinjowasp},
cross-site scripting~\cite{wasp}, 
and prototype pollution~\cite{prototypepoll} are all
examples of information flow vulnerabilities. A common strategy to defend against such violations
is to use {\em sanitizers} and {\em guards} in the code to block such bad flows.
Sanitizers and guards ensure that only safe information reaches the sensitive sinks. Information flow safety is a non-local property and both checking and fixing violations require non-local analysis. 

\definecolor{dIns}{RGB}{227,255,237}
\definecolor{dInsP}{RGB}{199,255,217}
\definecolor{dInsPP}{RGB}{163,242,190}
\newcommand{\cbox}[2]{\adjustbox{margin=0pt 1.9pt, bgcolor=#1}{#2}}
\newcommand{\cmbox}[2]{\adjustbox{margin=0pt 2.4pt, bgcolor=#1, margin=0pt -2.4pt}{#2}}
\newcommand{\cmboxp}[2]{\adjustbox{margin=0pt 2.4pt, bgcolor=#1, margin=0.7pt -0.7pt, cframe=red, margin=0pt -2.4pt}{#2}}
\newcommand{\INS}[1]{\cbox{dIns}{\textbf{+}}\cbox{dIns}{ #1}}
\newcommand{\mINS}[1]{\cbox{dIns}{#1}}
\newcommand{\INSp}[1]{\cbox{dInsPP}{+ #1}}
\newcommand{\mINSp}[1]{\cbox{dInsPP}{#1}}
\newcommand{\yINS}[1]{\cbox{yellow}{#1}}

\begin{figure}[!t]
    \centering
    \begin{subfigure}{0.5\textwidth}
    \begin{lstlisting}[autogobble,basicstyle=\scriptsize]
    var actions = new Map(); 
    loadActions(actions);
    app.get('/run', (req, res) => {
      var action = actions.get(req.action);
      if (action && <@\yINS{typeof action === 'function'}@>){<@\label{lst:line:witness-1-line}@>
        action(req.inp);
      }
    }
    \end{lstlisting}
    \caption{Safe program I for \unvalidatedtype vulnerability}\label{fig:static-witnessing-1}
    \end{subfigure}\hfil%
    \begin{subfigure}{0.5\textwidth}
    \begin{lstlisting}[autogobble,basicstyle=\scriptsize]
    var actions = new Map(); 
    loadActions(actions);
    app.get('/run', (req, res) => {
      var action = actions.get(req.action);
      if (<@\yINS{typeof action !== 'function'}@>)<@\label{lst:line:witness-2-line}@>
        return;
      action(req.inp);
    }
    \end{lstlisting}
    \caption{Safe program II for \unvalidatedtype vulnerability}\label{fig:static-witnessing-2}
    \end{subfigure}\hfil\\%
    \begin{subfigure}{0.5\textwidth}
    \begin{lstlisting}[escapechar=@,autogobble,basicstyle=\scriptsize]
    var actions = new Map(); 
    loadActions(actions);
    app.get('/run', (req, res) => {
      var action = actions.get(req.action);
      @\yINS{(typeof action === 'function')}@ && action(req.inp);@\label{lst:line:witness-3-line}@
    }
    \end{lstlisting}
    \caption{Safe program III for \unvalidatedtype vulnerability}\label{fig:static-witnessing-3}
    \end{subfigure}\hfil%
    \begin{subfigure}{0.5\textwidth}
    \begin{lstlisting}[escapechar=@,autogobble,basicstyle=\scriptsize]
    var actions = new Map(); 
    loadActions(actions);
    app.get('/run', (req, res) => {
      var action = actions.get(req.action);
      action(req.inp);        
    }
    \end{lstlisting}
    \caption{Unsafe program for \unvalidatedtype vulnerability}\label{fig:static-analysis}
    \end{subfigure}\hfil%
    \caption{\textit{\Safeprogs} that will be detected by \sawitness in (a), (b), and (c). The witnesses making programs safe are highlighted in yellow. (d) depicts an unsafe program will be detected by typical \sa.}
    \label{fig:static-analysis-vs-witnessing}
\end{figure}

We introduce \textbf{\sawitnessfull}, a new technique to collect a high-quality paired dataset of unsafe and safe programs. Instead of using the \sa tool to flag programs with violations, we use the internal information captured by the \sa\ tool on programs that satisfy the property, and identify the {\em reason} why the property is satisfied as a {\em witness}. In the case of information flow safety properties, these witnesses are usually sanitizers and guards in programs, which break the flow between untrusted sources and a trusted sinks.
By identifying and removing the witness, we introduce a violation and convert the safe program to an unsafe program, enabling us to construct safe-unsafe program pairs from programs that satisfy the property.

\lstMakeShortInline[columns=fixed]@
Consider the code snippets shown in Figure~\ref{fig:static-analysis-vs-witnessing}. In these code snippets, the value of @req@ comes from an untrusted source, and executing @action@ (which is dynamically derived from @actions@ map using @req.action@) can result in attacks. Programs (a), (b), and (c) are all examples of safe programs that satisfy the property, and the witnesses are highlighted. Program (d) is an unsafe program. By removing the witnesses, we transform each of the safe programs into unsafe variants and use these variants to construct safe-unsafe program pairs for the data collection phase.

For the strategy learning step, we use the paired set of programs as a training set and synthesize repair-strategies in a domain-specific-language (\dsl). We build on prior work in synthesizing program transformations from paired programs~\cite{rolim2017learning,zhang2022overwatch,bavishi2019phoenix}. However, information flow vulnerabilities are non-local, and repairing them is beyond the scope of previous work. Therefore, we design both a novel \dsl and a strategy learning algorithm that is able to learn strategies for such non-local repairs.
Consider the example shown in Figure~\ref{fig:vulnerabilty-example1}. The program in Figure~\ref{fig:unsafememberex} is vulnerable since there is a flow from the untrusted @JSON.Parse@ statement in line 12 to the trusted method invocation in line 6 without an intervening sanitizer. The repair here involves the introduction of the guard @if (handlers.hasOwnProperty(data.id))@ in line 5, as shown in Figure~\ref{fig:safememberex}. Learning such a repair involves (1) learning the location to introduce the repair (which is line 5 in this case), (2) learning the template of the guard that needs to be introduced (which is of the form @REF1.hasOwnProperty(REF2)@, where @REF1@ and @REF2@ are references, and (3) learning that @REF1@ and @REF2@ need to be materialized to @handlers@ and @data.id@ based on the given program. Our \dsl and strategy-learning algorithms (described in Section~\ref{sec:impl}) are novel, and are able to learn such intricate and non-local repairs, which involve analyzing both control and dataflows in the program.
\lstDeleteShortInline@

Prior works in static repair~\cite{bader2019getafix,bavishi2019phoenix,rolim2017learning,vurle,exampleBasedJava,sonarcube,revisar,genesis} make  simplifying assumptions. First, they assume the availability of  paired unsafe-safe program versions from version control. 
Second, the scope of repairs they consider are typically local.
Third, they consider statically typed languages like Java.
We focus on information flow vulnerabilities, which are inherently non-local, and we consider \js, which lacks static types. Moreover, we don't make the assumption that we have access to pairs of unsafe and safe programs.   
Instead, a single static snapshot of source-code repositories is sufficient for our technique.
Other approaches~\cite{hypergi,ACS:ICSE2017} in automatic program repair (APR) use program execution on vulnerability-causing examples which do not apply to repairing static 
vulnerabilities.
Crafting repair templates manually~\cite{senx,FootPatch,cdrep,bovinspector} is also challenging given the semantic nature of repairs.

We implemented the above approach consisting of static-analysis witnessing and strategy learning steps in a system \sysname.  In Section 6, we present an empirical evaluation of \sysname with code from several hundred {\sc JavaScript} open source repositories on Github. We use \codeql~\cite{codeql} as our \sa tool, and consider two specific instances of information flow vulnerabilities:
(1) unvalidated dynamic call (UDC), and (2) cross-site scripting (XSS).
For both these instances, we use static analysis witnessing and witness removal to generate unsafe-safe pairs, and learn repair strategies from this data. Then we evaluate the effectiveness of the learned repair strategies on all the violations detected by \codeql in these repositories. Thus, our training set is generated from correct coding patterns (which is that static analysis witnessing uses), and our validation set is the set of violations in these repositories (which are disjoint from the training set). We find that \sysname is able to correctly repair: (1)  {\bf 310} vulnerabilities with a success rate of {\bf 93.94\%}, in the case of unvalidated dynamic call, and (2) {\bf 617} vulnerabilities with a success rate of {\bf 91.82\%}, in the case of cross-site scripting. We compare the results with two neural baselines, namely one obtained by finetuning {\sc CodeT5}~\cite{Wang2021CodeT5IU} and the other obtained by few-shot prompting {\sc Codex}\cite{Codex} with the same training 
set as \sysname. We find that \sysname\ outperforms both these neural baselines.

\medskip
\noindent
To summarize, our main contributions are:
\begin{itemize}
\item
A new approach called static-analysis witnessing to produce unsafe-safe code pairs from programs that satisfy a property, and
\item
A novel DSL and strategy learning algorithm to learn non-local repair strategies from paired data sets (such as the ones generated from static analysis witnessing, or other approaches)
\end{itemize}
We present an implementation of the approaches in a tool, \sysname.  Our empirical results show that \sysname is able to correctly repair hundreds of violations in open source {\sc JavaScript} repositories, while outperforming neural baselines. We will publicly release the datasets used in our evaluation (Section~\ref{sec:experiments}).

\section{Background And Overview}
\label{sec:background}
\begin{table}[!t]
    \centering
    \begin{tabular}{|p{0.17\linewidth} | p{0.80\linewidth}|}
       \hline
       \textbf{Term} & \textbf{Definition} \\
       \hline
       source & a variable whose value is directly set by an (untrusted) user \\
       tainted variable & a variable whose value is derived from a source variable and is controllable by an (untrusted) user; the value of such a variable is called tainted value \\
       sink & a program execution performing a security-critical operation using an input \\
       sanitizer & function that takes the tainted variable as input and removes the taint\\
       guard & a check performed on the tainted values to block execution on malicious inputs \\
       \hline
    \end{tabular}
    \caption{\taintprop vulnerability terminologies and definitions}
    \label{tab:taint-terminologies}
\end{table}


\subsection{Problem Background and Motivating Examples}

\newcommand{\mynode}[2]{\tikz[remember picture]\node[#2,shape=coordinate](#1){};}
\definecolor{edgecolor}{HTML}{008080}

\newcommand{\wrapbox}[2]{\adjustbox{margin=1.5pt 1.5pt 1.5pt 3pt, bgcolor=white, frame=1pt, cframe=#1, color=#1}{#2}}

\newcommand{\wrapboxintext}[2]{\adjustbox{margin=1.5pt 1.5pt, bgcolor=white, frame=1pt, cframe=#1, color=#1}{#2}}

\tikzset{
  ex1arrow/.style={, arrows={-{Stealth[length=2mm, width=1.5mm]}}, line width=0.2ex, edgecolor}
}

\begin{figure}
\centering
\begin{subfigure}{0.48\textwidth}
	\begin{lstlisting}
handlers["run"] = function (data<@\mynode{n5}{above right=0.5ex}\label{lst:line:handlers-run}@>){
    . . .
    let foo<@\mynode{n7}{above left=0.9ex}\label{lst:line:callerId-index}@> = handlers[data.<@\mynode{n6}{above left=0.5ex}@>id];
    . . .
    
    <@\mynode{n8}{above left=0.1ex}\label{lst:line:callerId-sink}@><@\wrapbox{red}{foo}@>(data);

}<@\label{lst:line:handlers-run-end}@>
. . .
var commHandler = function (<@\wrapbox{orange}{event}@><@\mynode{n1}{above right=0.5 ex}\label{lst:line:commHandler-source}@>){
   . . .
   var <@\mynode{n3}{above left=0.8ex}@>data = JSON.parse(<@\mynode{n2}{above=1.5ex}@>event.data);
   handlers["run"](data<@\mynode{n4}{above right=1ex}@>);
}\end{lstlisting}
	\begin{tikzpicture}[remember picture, overlay,
    every edge/.append style = { ->, thick, >=stealth, dashed, line width = 2pt }]

      \draw[ex1arrow] (n1) to[bend left=15]  node[circle,sloped,inner sep=1pt,draw,pos=0.6,above=1pt] {\scriptsize 1} (n2);
      \draw[ex1arrow] (n2) to[bend right=25] node [circle,sloped,inner sep=1pt,draw,midway,above=1pt] {\scriptsize 2} (n3) ;
      \draw[ex1arrow] (n3) to[bend right=45] node [circle,sloped,inner sep=1pt,draw,midway,below=1pt] {\scriptsize 3} (n4) ;
      \draw[ex1arrow] (n4) to[bend right=90,looseness=1.7] node [circle,sloped,inner sep=1pt,draw,midway,above=10pt,right=15pt] {\scriptsize 4} (n5);
      \draw[ex1arrow] (n5) to[bend left=40] node [circle,sloped,inner sep=1pt,draw,pos=0.5,left=1pt,above=1pt] {\scriptsize 5} (n6);
      \draw[ex1arrow] (n6) to[bend right=50] node [circle,sloped,inner sep=1pt,draw,midway,below=0.8pt] {\scriptsize 6} (n7);
      \draw[ex1arrow] (n7) to[bend right=90,looseness=1.8] node [circle,sloped,inner sep=1pt,draw,pos=0.8,left=1pt,above=1pt] {\scriptsize 7} (n8);
\end{tikzpicture}
\caption{Example for vulnerable program}
\label{fig:unsafememberex}
\end{subfigure}\hfil%
\begin{subfigure}{0.48\textwidth}
	\begin{lstlisting}
handlers["run"] = function (data<@\mynode{n5}{above right=0.5ex}@>){
    . . .
    let foo = handlers[data.id];
    . . .
<@\INSp{\ \ \ \ if handlers.hasOwnProperty(data.id)\{}\label{lst:line:fix-start}@>
<@\mINS{\ \ \ \ \ \ \ \ \ \ foo(data);}@>
<@\INSp{\ \ \ \ \}}@><@\label{lst:line:fix-end}@>
}
. . .
var commHandler = function (event<@\mynode{n1}{above right=0.5ex}@>){
   . . .
   var <@\mynode{n3}{above left=0.4ex}@>data = JSON.parse(<@\mynode{n2}{above left=0.5ex}@>event.data);
   handlers["run"](data<@\mynode{n4}{above right=0.5ex}@>);
}\end{lstlisting}
	\begin{tikzpicture}[remember picture, overlay,
    every edge/.append style = { ->, thick, >=stealth, dashed, line width = 1pt }]

\end{tikzpicture}
\caption{Example for fixed safe program}
\label{fig:safememberex}
\end{subfigure}

\caption{Example and corresponding fix for the \unvalidatedmembership vulnerability.}
\label{fig:vulnerabilty-example1}

\end{figure}
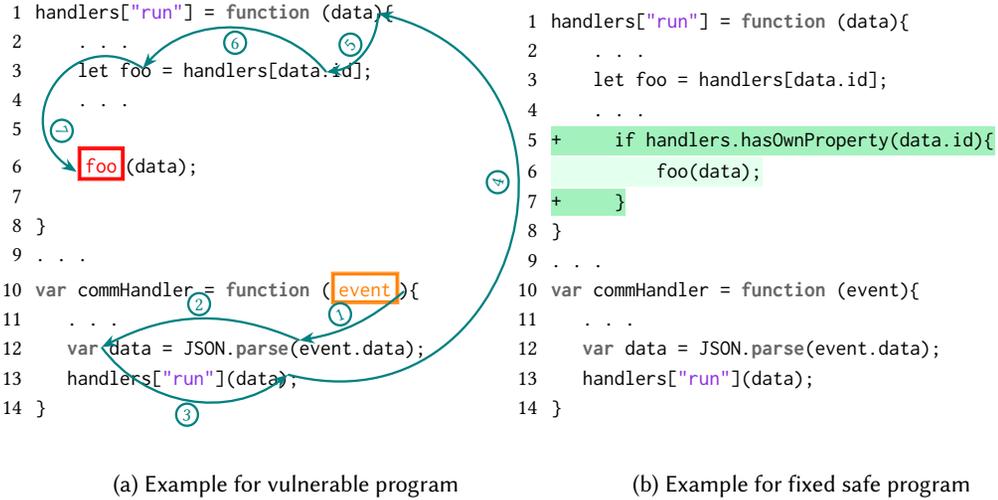

\lstMakeShortInline[columns=fixed]@
In the previous section we defined information flow vulnerabilities as flow of information from an untrusted source to a trusted or sensitive sink without appropriate sanitization.
In Table~\ref{tab:taint-terminologies} we define these terms more precisely. 
\sa tools find these vulnerabilities by detecting sources and sinks, and then performing a dataflow analysis between them.
We call a program unsafe or safe depending on whether \sa detects a violation or not. Further, we call a program a  {\em \safeprog} if it is safe because a {\em witness}, i.e., a sanitizer or a guard, blocks the flow between a source and a sink. Note that a safe program is not necessarily a witnessing-safe program. In particular, a program that does not have any sources or sinks is vacuously safe, but not a witnessing-safe program.
Next, we give two simplified examples with violations and show how one can fix them. 

\noindent \textbf{Example 1.} Figure~\ref{fig:unsafememberex} is an unsafe program containing the \unvalidatedmembership vulnerability where an untrusted user input is used as a key to index into a record of functions. A malicious user can exploit this vulnerability by passing missing keys or keys defined by parent or base classes like @"__proto__"@ or @"constructor"@.
Here, the vulnerability arises from the flow between the source @event@ and the unvalidated function call on line 6. The @commHandler@ function in Line~\ref{lst:line:commHandler-source} takes in a user input in the form of @event@ variable, thus making @event@ the source (highlighted in orange). This source variable now propagates taint across expressions in the program, as depicted by the dataflow edges 1-7 (in cyan). Specifically @event.data@, @data@, and @data.id@ are all tainted expressions. Notice that the \taintprop happens across method boundaries through the @handlers["run"]@ function in Line~\ref{lst:line:handlers-run}. In Line~\ref{lst:line:callerId-index}, the @handlers@ object is dynamically indexed with the @data.id@ tainted-variable and the indexed value is called as a function in Line~\ref{lst:line:callerId-sink} 
without checking whether the record  @handlers@ has a function with key @data.id@ as its ``own'' property\footnote{\url{https://developer.mozilla.org/en-US/docs/Web/JavaScript/Reference/Global\_Objects/Object/hasOwnProperty}}. 

Figure~\ref{fig:safememberex} depicts a  corresponding safe program with the transformation highlighted in green. The vulnerability is fixed by replacing the statement containing the sink, \wrapbox{red}{foo(data)} with the if-statement (guard) between Lines~\ref{lst:line:fix-start} and ~\ref{lst:line:fix-end}. This guard blocks the execution of the sink on malicious user inputs. Specifically, @handlers.hasOwnProperty(data.id)@ (Line ~\ref{lst:line:fix-start}) checks whether @data.id@ is indeed an \textit{own property} of the object and blocks the execution of the sink otherwise. Notice that the variables in the guard  (@handlers@ and @data.id@) do not syntactically appear in the sink statement (\wrapboxintext{red}{foo(data)}). Hence,  the repair strategy must use  non-local dataflow information to repair.

\noindent \textbf{Example 2.} Figure~\ref{fig:unsafexss}
is an unsafe program with the cross-site scripting (\xss for short) vulnerability where untrusted user-input flows to an HTML response~\cite{xsscodeql,wasp}. A user can create a request where the @req.id@ attribute contains malicious \js code; this code will get executed on the application side making the program unsafe. The
@requestListener@ HTTP server handler on Line~\ref{lst:line:requestlistener} reads the @userId@ input in Line~\ref{lst:line:userId}. This @userId@ is used internally to handle the request as needed and then concatenated with a prefix into the @message@ variable in Line~\ref{lst:line:concat}. Finally, the tainted @message@ variable is sent inside the HTTP response sink. 

\begin{figure}
\centering
\begin{subfigure}{0.48\textwidth}
	\begin{lstlisting}[basicstyle=\footnotesize]
const requestListener = function (<@\mynode{n0}{above left=0.5ex}@><@\wrapbox{orange}{req}@>, res) {<@\label{lst:line:requestlistener}@>
  <@{\Large $\cdots$}@>
  var userId<@\mynode{n2}{above right=0.5ex}\label{lst:line:userId}@> = req.id<@\mynode{n1}{above right=0.9ex}@>;

  serveRequest(userId<@\mynode{n3}{above right=0.9ex}@>);
  <@{\Large $\cdots$}@>
  var message<@\mynode{n5}{above right=0.3ex} = "Served user " + userId<@\mynode{n4}{above right=0.9ex}\label{lst:line:concat}@>;
  <@{\Large $\cdots$}@>
  res.send({"msg": <@\mynode{n6}{above left=0.1ex}\wrapbox{red}{message}\label{lst:line:response}@>});
};<@\label{lst:line:requestlistener-end}@>
\end{lstlisting}
	\begin{tikzpicture}[remember picture, overlay,
    every edge/.append style = { ->, thick, >=stealth, dashed, line width = 2pt }]

      \draw[ex1arrow] (n0) to[bend left=30]  node[circle,sloped,inner sep=1pt,draw,pos=0.6,above=1pt] {\scriptsize 1} (n1);
      \draw[ex1arrow] (n1) to[bend right=60]  node[circle,sloped,inner sep=1pt,draw,pos=0.6,above=1pt] {\scriptsize 2} (n2);
      \draw[ex1arrow] (n2) to[bend left=10] node [circle,sloped,inner sep=1pt,draw,pos=0.8,above=1pt] {\scriptsize 3} (n3) ;
      \draw[ex1arrow] (n3) to[bend left=45] node [circle,sloped,inner sep=1pt,draw,midway,below=1pt] {\scriptsize 4} (n4) ;
      \draw[ex1arrow] (n4) to[bend right=35] node [circle,sloped,inner sep=1pt,draw,midway,below=1pt] {\scriptsize 5} (n5) ;
      \draw[ex1arrow] (n5) to[bend right=5] node [circle,sloped,inner sep=1pt,draw,pos=0.3,below=1pt] {\scriptsize 6} (n6) ;
\end{tikzpicture}
\caption{Example for vulnerable program}
\label{fig:unsafexss}
\end{subfigure}\hfil%
\begin{subfigure}{0.48\textwidth}
	\begin{lstlisting}[basicstyle=\footnotesize]
const requestListener = function (req, res) {
  <@{\Large $\cdots$}@>
  var userId = req.id;
<@\INSp{\ userId = {\color{darkgray} \textbf{escape}}(userId);}\label{lst:line:escape-fix}@>
  serveRequest(userId);
  <@{\Large $\cdots$}@>
  var message = "Served user " + userId;
  <@{\Large $\cdots$}@>
  res.send({"msg": message});
};
\end{lstlisting}
	\begin{tikzpicture}[remember picture, overlay,
    every edge/.append style = { ->, thick, >=stealth, dashed, line width = 1pt }]

\end{tikzpicture}
\caption{Example for fixed safe program}
\label{fig:safexss}
\end{subfigure}

\caption{Example and corresponding fix for the \xss vulnerability}
\label{fig:vulnerabilty-example2}

\end{figure}

Figure~\ref{fig:safexss} depicts a corresponding safe program with the transformation highlighted in green. The vulnerability is fixed by inserting a new statement containing the sanitizer. It removes the taint from a variable and makes execution of the sink on the variable safe. Specifically, the @escape@ function in Line~\ref{lst:line:escape-fix} sanitizes the @userId@ variable and removes the taint.
Note that  this transformation is applied at an intermediate location between the source and the sink.

In this paper, we are interested in automatically repairing  \taintprop vulnerabilities by introducing sanitizers and guards. 
We first use a snapshot of a training code base to learn repair strategies. Next, given
an input program with a violation, we provide a high-level overview of how these learned strategies repair the input.


\subsection{Applying repair strategies}
Given an AST of an unsafe program annotated with sources, sinks, and vulnerable flows, our repair strategies follow a two-step process: find the \textit{edit-location} and then apply an \textit{edit-operation}, where:
\begin{enumerate}
    \item the edit-location is an \astree-node where the edit occurs, and
    \item the edit-operation is a tree-edit operation at the edit-location. Since these vulnerabilities are fixed by introducing sanitizers or guards in programs, we support inserting a child and replacing a child with another tree as the edit operations. 
\end{enumerate}

\textbf{Example 1.} The repair in Figure~\ref{fig:vulnerabilty-example1} introduces an if-statement with @handlers.hasOwnProperty(data.id)@ guard that makes the program safe. 
Thus the edit-location is the \astree-node corresponding to the block statement between Lines~\ref{lst:line:handlers-run} and ~\ref{lst:line:handlers-run-end}. The edit-operation replaces the child containing the sink \astree-node with the if-statement.

\textbf{Example 2.} The repair in Figure~\ref{fig:vulnerabilty-example2} inserts an assignment statement at Line~\ref{lst:line:escape-fix} with the @escape@ sanitizer to make the program safe.  Thus the edit-location is the \astree corresponding to the block statement between Lines~\ref{lst:line:requestlistener} and ~\ref{lst:line:requestlistener-end}. The edit-operation inserts the assignment statement as an additional child to this block. 

The repair strategies have two components: (1) an edit localization component, which predicts the edit-location, and (2) an edit operation component, which constructs the input-program-specific edit that needs to be applied at the edit-location. We describe these components below.

\medskip
\noindent \textbf{Edit Localization}.
This component of the repair strategy encodes a path to the edit location starting from some known location in the program. In the case of information flow violations, the \sa tool outputs the locations of the source and sink, and our repair strategy makes use of these anchors. An example encoding of reaching the edit location from the location of the source is as follows: \textit{traverse data flow semantic edges in the annotated AST from the source along a path till a function call is encountered. Then, traverse syntactic AST edges from this call node to find the innermost block statement which is an ancestor to the function call. That block statement is the edit-location}. 
In Section 4 (Figure 9), we show a DSL for representing repair strategies. The above encoding is represented in this DSL using a \kleeneedge that allows crossing multiple edges of a particular \textit{kind} until reaching a stopping-location. 
In Figure~\ref{fig:vulnerabilty-example1}, this localization component follows the semantic edges 1-7 from \wrapboxintext{orange}{event} to reach the \wrapboxintext{red}{foo} call in Line~\ref{lst:line:callerId-sink}. Then, we traverse  the syntactic parent of the @foo@ call multiple times  to reach the block statement of lines 1--8. Abstracting over the number of edges using \kleeneedge enables the localization component to generalize over other programs, which could be syntactic variations of this example. 
We learn these edge-types (syntactic or semantic) and stopping-locations (function call and block statement) from the training data.

\medskip
\noindent \textbf{Edit Operation.}
This component has abstract programs that are instantiated to ASTs using the input program. For example, a repair strategy can have an abstract guard @REF1.hasOwnProperty(REF2){REF3}@, where @REF1@ @REF2@ @REF3@ are materialized with AST nodes that can be obtained by traversing {\em reference paths} of the input program. 
For Figure~\ref{fig:vulnerabilty-example1},
instantiating this abstract guard with the unsafe program provides the guard @handlers.hasOwnProperty(data.id){foo(data)}@. Here, @REF2@ (which materializes into @data.id@) is found by traversing the {\em reference path} containing @semantic-parent@ edges from \textit{semantic-location}\footnote{Semantic Location is the node on the edit-path at the boundary between semantic-edges and syntactic-edges defined in Section~\ref{sec:learning}}  \wrapboxintext{red}{foo}. Similarly, @REF1@ and @REF3@ materialize into @handlers@ and @foo(data);@ respectively by traversing reference paths associated with them.

\lstDeleteShortInline@

\section{Data Collection}
\label{sec:data}
\lstMakeShortInline[columns=fixed]@
\begin{figure}[h] 
\centering
    \begin{subfigure}[t]{0.53\textwidth}
\usetikzlibrary{automata,shapes,shapes.geometric,arrows,fit,calc}
\tikzset{->,elliptic state/.style={draw,ellipse}}
\begin{tikzpicture}[shorten >=1pt, scale = 0.5, transform shape]

\node[elliptic state, align=center, blue] (A) [] {\{$\dots$\}\\\blockstmt};

\node[elliptic state, align=center, left of = A, xshift=-2.5cm, yshift=-1.6cm] (B) [] {let foo = $\dots$\\\declexpr};
\node[elliptic state, align=center, right of = A, xshift=3.5cm, yshift=-1.6cm] (C) [] {foo(data);\\\expr};

\node[elliptic state, align=center, below of = B, xshift=0cm, yshift=-1.5cm] (D) [] {foo = $\dots$\\\declarator};
\node[elliptic state, align=center, below of = C, xshift=0cm, yshift=-1.5cm] (E) [] {foo(data)\\\callexpr};

\node[elliptic state, align=center, left of = D, xshift=-1cm, yshift=-2.2cm] (F) [] {foo\\\vardecl};
\node[elliptic state, align=center, right of = D, xshift=1cm, yshift=-2.2cm] (G) [] {handlers[data.id]\\\indexexpr};
\node[elliptic state, align=center, left of = E, xshift=-1cm, yshift=-2.2cm, very thick, red] (H) [] {foo\\\varexpr};
\node[elliptic state, align=center, right of = E, xshift=1cm, yshift=-2.2cm] (I) [] {data\\\varexpr};

\coordinate[below left of=G, yshift=-1cm, xshift=-1.5cm] (G1);
\coordinate[below right of=G, yshift=-1cm, xshift=1.5cm] (G2);

\coordinate[left of=A, yshift=-0.3cm, xshift=-4cm] (A0);
\coordinate[right of=A, yshift=-0.3cm, xshift=4cm] (Ae);
\coordinate[left of=A, yshift=-0.7cm, xshift=-3.5cm] (A01);
\coordinate[right of=A, yshift=-0.7cm, xshift=3.5cm] (Ae1);

\coordinate[left of=A, yshift=-3cm, xshift=-0.4cm] (A1);
\coordinate[right of=A, yshift=-3cm, xshift=0.4cm] (A2);
\coordinate[left of=A, yshift=-3cm, xshift=1cm] (A3);


\node[elliptic state, align=center, below of = I, xshift=-3cm, yshift=-2cm, very thick, orange] (S) {event\\\varexpr};
\node[elliptic state, align=center, left of = S, xshift=-4cm, yshift=0cm] (Smid) {data\\\varexpr};

\draw (A) edge[] node [elliptic state,sloped,inner sep=1pt,draw,pos=0.5,right=1pt,below=1pt] {ch:7} (B);
\draw (A) edge[] node [elliptic state,sloped,inner sep=1pt,draw,pos=0.5,right=1pt,below=1pt] {ch:13} (C);

\draw (B) edge[] node [elliptic state,sloped,inner sep=1pt,draw,pos=0.5,right=1pt,above=1pt] {ch:0} (D);
\draw (C) edge[] node [elliptic state,sloped,inner sep=1pt,draw,pos=0.5,right=1pt,above=1pt] {ch:0} (E);

\draw (D) edge[] node [elliptic state,sloped,inner sep=1pt,draw,pos=0.5,right=1pt,above=1pt] {ch:0} (F);
\draw (D) edge[] node [elliptic state,sloped,inner sep=1pt,draw,pos=0.5,right=1pt,above=1pt] {ch:1} (G);
\draw (E) edge[] node [elliptic state,sloped,inner sep=1pt,draw,pos=0.5,right=1pt,above=1pt] {ch:0} (H);
\draw (E) edge[] node [elliptic state,sloped,inner sep=1pt,draw,pos=0.5,right=1pt,above=1pt] {ch:1} (I);

\draw (A) edge[] node [elliptic state,sloped,inner sep=1pt,draw,pos=0.5,right=1pt,above=1pt] {ch:0} (A0);
\draw (A) edge[] node [elliptic state,sloped,inner sep=1pt,draw,pos=0.5,right=1pt,above=1pt] {ch:20} (Ae);

\draw (A) edge[] node [elliptic state,sloped,inner sep=1pt,draw,pos=0.5,right=1pt,below=1pt] {ch:8} (A1);
\draw (A) edge[] node [elliptic state,sloped,inner sep=1pt,draw,pos=0.5,right=1pt,below=1pt] {ch:12} (A2);
\draw (A) edge[] (A3);
\draw (G) edge[] node [elliptic state,sloped,inner sep=1pt,draw,pos=0.5,right=1pt,above=1pt] {ch:0} (G1);
\draw (G) edge[] node [elliptic state,sloped,inner sep=1pt,draw,pos=0.5,right=1pt,above=1pt] {ch:1} (G2);

\draw (A) edge[] (A01);
\draw (A) edge[] (Ae1);


\draw (S) edge[bend left=0, left, edgecolor] node [circle,sloped,inner sep=1pt,draw,pos=0.5,right=1pt,above=1pt] {1-3} node [sloped,pos=0.5,left=0.5pt,below=0.5pt] {semantic} (Smid);
\draw (Smid) edge[bend left=0, left, edgecolor] node [circle,sloped,inner sep=1pt,draw,pos=0.5,right=0.5pt,above=0.5pt] {4-5} node [sloped,pos=0.5,left=1pt,below=1pt] {semantic} (G);
\draw (G) edge [edgecolor] node [circle,sloped,inner sep=1pt,draw,pos=0.5,right=1pt,above=1pt] {6} node [sloped,pos=0.5,left=0.5pt,below=0.5pt] {semantic} (F);
\draw (F) edge [edgecolor,bend right = 53] node [circle,sloped,inner sep=1pt,draw,pos=0.9,right=1pt,above=1pt] {7} node [sloped,pos=0.9,left=0.5pt,below=0.5pt] {semantic} (H);
\end{tikzpicture}
\caption{\pdg representation for the unsafe program in Figure~\ref{fig:unsafememberex} with the source in orange and the sink in red}
\label{fig:example1-pdg}
\end{subfigure}\hfil%
\begin{subfigure}[t]{0.42\textwidth}
\usetikzlibrary{automata,shapes,shapes.geometric,arrows,fit,calc}
\tikzset{->,elliptic state/.style={draw,ellipse}}
\begin{tikzpicture}[shorten >=1pt, scale = 0.5, transform shape]

\node[elliptic state, align=center] (A) {if hand...ata) \{..\}\\\ifstmt};
\node[elliptic state, align=center, below of = A, xshift=-3cm, yshift=-1.5cm] (B) {handlers.hasOwnProperty(data.id)\\\callexpr};
\node[elliptic state, align=center, below of = A, xshift=2.5cm, yshift=-1.5cm] (C) {\{..\}\\\blockstmt};
\node[circle, fill=black, radius=1cm, align=center, below of = C, xshift=1.6cm, yshift=-2cm] (CHole) {};
\node[elliptic state, align=center, below of = B, xshift=-1cm, yshift=-2cm] (D) {handlers.hasOwnProperty\\\dotexpr};
\node[elliptic state, align=center, below of = B, xshift=5cm, yshift=-2cm] (E) {data.id\\\dotexpr};
\node[elliptic state, align=center, below of = D, xshift=-2cm, yshift=-3cm] (F) {handlers\\\varexpr};
\node[elliptic state, align=center, below of = D, xshift=3cm, yshift=-3cm] (G) {hasOwnProperty\\\labelexpr};

\node[elliptic state, align=center, below of = E, xshift=1.5cm, yshift=-1.5cm] (H) {data\\\varexpr};
\node[elliptic state, align=center, below of = E, xshift=-1.5cm, yshift=-1.5cm] (I) {id\\\varexpr};

\draw (A) edge[] node [elliptic state,sloped,inner sep=1pt,draw,pos=0.5,right=1pt,above=1pt] {ch:0} (B);
\draw (A) edge[] node [elliptic state,sloped,inner sep=1pt,draw,pos=0.5,right=1pt,above=1pt] {ch:1} (C);
\draw (C) edge[dashed] node [elliptic state,sloped,inner sep=1pt,draw,pos=0.5,right=1pt,above=1pt] {ch:0} (CHole);
\draw (B) edge[] node [elliptic state,sloped,inner sep=1pt,draw,pos=0.5,right=1pt,above=1pt] {ch:0} (D);
\draw (B) edge[] node [elliptic state,sloped,inner sep=1pt,draw,pos=0.5,right=1pt,above=1pt] {ch:0} (E);
\draw (D) edge[] node [elliptic state,sloped,inner sep=1pt,draw,pos=0.5,right=1pt,above=1pt] {ch:0} (F);
\draw (D) edge[] node [elliptic state,sloped,inner sep=1pt,draw,pos=0.5,right=1pt,above=1pt] {ch:1} (G);
\draw (E) edge[] node [elliptic state,sloped,inner sep=1pt,draw,pos=0.5,right=1pt,above=1pt] {ch:0} (H);
\draw (E) edge[] node [elliptic state,sloped,inner sep=1pt,draw,pos=0.5,right=1pt,above=1pt] {ch:1} (I);

\end{tikzpicture}
\caption{\astree representation for the editprog corresponding to the unsafe program in Figure~\ref{fig:unsafememberex}}
\label{fig:example1-editprog}
\end{subfigure}

\caption{Example of \pdg and \astree of editprog }
\end{figure}
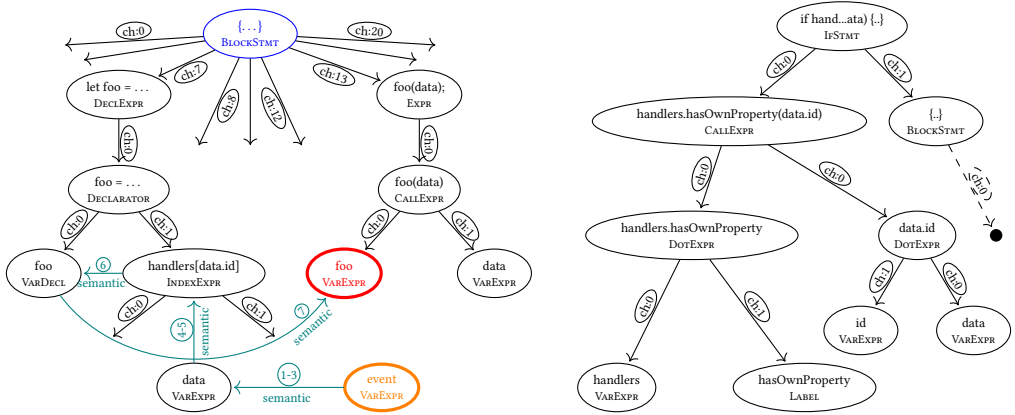
\lstDeleteShortInline@

In this section, we describe how we collect examples for learning repair strategies without any version-controlled data. Specifically, we first detect \safeprogs and corresponding witnesses using \sawitnessfull (witnesses are sanitizers and guards that protect from vulnerabilities)  in Section~\ref{subsec:sa-witness}. Using these witness annotations, we generate unsafe programs and \textit{edits} from the \safeprog using a \textbf{witness-removal} step (Section ~\ref{subsec:witness-removal}). In the following, we define terminology for the \astree  data-structure we operate on.

\astree refers to the abstract syntax tree representation of programs, augmented with data flow edges and annotations for sources, sinks, sanitizers, guards, witnesses etc. 
An \astree is a five-tuple 
$\langle \mathcal{N},\mathcal{V},\mathcal{T},\mathcal{E}, \mathcal{A} \rangle$, where:
\begin{enumerate}
\item
$\mathcal{N}=\{\mathit{id}_0,\ldots\mathit{id}_n\}$  is a set of nodes, where  $\mathit{id_i}\in\mathbb{N}$ for 
$ 0 \leq i \leq n$.
\item
$\mathcal{V}$ is a map from nodes to program snippets
represented as strings. For a node $n$, we have that $\mathcal{V}(n)$ is a string representing the code snippet associated with $n$
\item
$\mathcal{T}$ is a map from nodes to their types defined by 
 \sa~\cite{codeqlast}. For example, \callexpr is the type of a node representing a function call, \indexexpr is the type of a node representing an array index, and \blockstmt is the type of a node representing a basic block of statements.
\item
$\mathcal{E}$ is a set of directed edges.
Each edge is of the form $(n_1,n_2,\edgetype,z)$, where
$n_1$ is a source node, $n_2$ is a target node, 
$\edgetype \in \{\T{SynParent}, \T{SynChild}, \T{SemParent},
\T{SemChild} \}$ denotes the relationship from 
$n_1$ to $n_2$, as one of syntactic parent, syntactic child, semantic parent or semantic child,
and $z\in\mathbb{Z}$ is the index of $n_2$ among $n_1's$ children if this edge is a child edge, and $-1$ if the edge is a parent edge. 
\item
$\mathcal{A}$ is a set of annotations associated with each node. The annotations are from the set $\{\T{source},
\T{sink},\T{sanitizer},\T{guard}$,\T{witness}\}. We also refer to annotations using predicates or relations. For instance, for a node $n$, if an annotation  $\T{source}$ is present, we say that
the predicate $\T{source}(n)$ is true.
\end{enumerate}

A {\em traversal} or a {\em path} in an \astree is a sequence of edges $e_0,\ldots,e_{i-1},e_i,\ldots ,e_k$ such that the target node of $e_{i-1}$ is also the source node of $e_i$, for all $i\in\{1,\ldots,k\}$. That is, $e_{i-1}$ is of the form $(\_,n,\_,\_)$ and $e_i$ is of the form $(n,\_,\_,\_,\_)$. The source node of $e_0$ is the source of this path and the target node of $e_k$ is the target of the path.

\lstMakeShortInline[columns=fixed]@
Figure~\ref{fig:example1-pdg} depicts a partial \pdg corresponding to the unsafe program in Figure~\ref{fig:unsafememberex}. Each oval corresponds to an \astree-node containing a type $\tau$ and an associated value. The dark edges denote the syntactic child edges. For example, the oval with value @foo(data)@ is an \astree-node with type \callexpr and has two children -- @foo@ and @data@, both with the type \varexpr. 
The semantic child edges are at the bottom in cyan. These edges correspond to the ones depicted in cyan in Figure ~\ref{fig:unsafememberex}. 
\lstDeleteShortInline@


If $\prog$ is an \pdg then
we use  $\prog.\mathtt{source}$ to denote the source node, $\prog.\mathtt{sink}$ to denote the sink node, and $\prog.\mathtt{witness}$ to denote the witness node.
If the program has several sources, sinks and sanitizers then we generate a separate \pdg for each $(\mathtt{source},\mathtt{witness},\mathtt{sink})$ triple.
For a node $n$, its syntactic parent is $n.\mathtt{parent}$, syntactic children are $n.\mathtt{children}$, semantic parent is $n.\mathtt{semparent}$, and semantic children are $n.\mathtt{semchildren}$.

\subsection{Static Analysis Witnessing}
\label{subsec:sa-witness}

\newcommand{\DMethodjudge}[1]{\texttt{#1(}\checknextarga}

\begin{figure}[t]
    \scriptsize{
    \begin{adjustbox}{width=\columnwidth}
    \begin{mathpar}

    \inferrule*[]
    {\DMethodjudge{SemChild}{n_1}{n_2}\\
    \neg\DMethodjudge{SanGuard}{n_1}\\
    \neg\DMethodjudge{SanGuard}{n_2}}
    {{\color{red} \DMethodjudge{SanGuardFree}{n_1}{n_2}}}

    \inferrule*[]
    {\DMethodjudge{Source}{n_1}\\
    \DMethodjudge{Sink}{n_2}\\
    {\color{red} \DMethodjudge{SanGuardFree*}{n_1}{n_2}}}
    {\DMethodjudge{\textbf{Vulnerability}}{n_1}{n_2}}

    \inferrule*[]
    {\DMethodjudge{SemChild*}{n_1}{n_3}\\
    \DMethodjudge{SemChild*}{n_3}{n_2}\\
    \DMethodjudge{SanGuard}{n_3}}
    {{\color{ForestGreen} \DMethodjudge{SanGuardInMid}{n_1}{n_3}{n_2}}}

    \inferrule*[]
    {\DMethodjudge{Source}{n_1}\\
    \DMethodjudge{Sink}{n_2}\\
    {\color{ForestGreen} \DMethodjudge{SanGuardInMid}{n_1}{n_3}{n_2}}}
    {\DMethodjudge{\textbf{Witness}}{n_1}{n_3}{n_2}}

    \end{mathpar}
    \end{adjustbox}
    }
    \caption{Judgement rules for \T{Vulnerability} and \T{Witness} relations}
    \label{fig:judgements}
    
\end{figure}

In this section, we show how to repurpose \sa tools to generate witnesses.
\sa tools perform dataflow analysis to check for rule-violations in programs. They use pattern matching to identify known sources, sinks, sanitizers, and guards. For commercial tools, these patterns are implemented (and continuously updated) manually by developers and encode this domain knowledge. Next, 
\sa checks if there exists a flow between a source and a sink that does not cross a sanitizer or guard. We capture this formally in Figure~\ref{fig:judgements} (top two rules), and explain the notation used in it below.

\sa tools encode domain knowledge about the vulnerability by annotating nodes as \T{Source}, \T{Sink}, \T{Sanitizer}, and \T{Guard}. 
So \DMethod{Source}{\I{n}}\ is true iff the node \I{n} is a \textit{source} node for a vulnerability. Next, \sa tools perform dataflow analysis by defining the relation \DMethod{SemChild}{$n_1$}{$n_2$}\ which is true iff there is a \taintpropedge between $n_1$ and $n_2$. Then the \DMethod{Vulnerability}{$n_1$}{$n_2$}\ relation can be defined as:
\begin{enumerate}
    \item $n_1$ and $n_2$ are source and sink nodes (\DMethod{Source}{$n_1$}\ and \DMethod{Sink}{$n_2$}\ are true)
    \item There exists a \textit{path} between $n_1$ and $n_2$ which is free of sanitizers or guards (\DMethod{SanGuardFree*}{$n_1$}{$n_2$}\ is true). A path is free of sanitizers and guards iff every \textit{edge} in the \textit{path} is free of sanitizers and guards. An edge between $n_1$ and $n_2$ is considered free of sanitizers and guards (\DMethod{SanGuardFree}{$n_1$}{$n_2$}\ is true) iff $(n_1, n_2, \_, \T{SemChild}) \in \mathcal{E}$ and neither of $n_1$ or $n_2$ is a sanitizer or a guard
\end{enumerate}

Here, we make the following observation - \emph{this domain knowledge present in these annotations and relations is helpful beyond just detecting vulnerabilities}. For instance, simply using the sanitizer relation allows us to query the different kinds of sanitizers domain experts have specified. We use this observation to discover \emph{\safeprogs} i.e., programs having a source, sink, and a sanitizer or guard that \textit{blocks} the \taintprop or, in simpler terms, make the program safe. In addition, we also detect the corresponding sanitizers or guards in the programs and refer to them as \textit{witnesses} because they serve as the evidence of making the program safe. We call this procedure \sawitnessfull (abbreviated as \sawitness). 
We define this as the \T{Witness} relation in Figure~\ref{fig:judgements} (bottom two rules). Specifically, \DMethod{Witness}{$n_1$}{$n_3$}{$n_2$}\ is defined as:
\begin{enumerate}
    \item $n_1$ and $n_2$ are source and sink nodes (\DMethod{Source}{$n_1$}\ and \DMethod{Sink}{$n_2$}\ are true)
    \item There exists a node $n_3$ such that it satisfies \DMethod{SanGuardInMid}{$n_1$}{$n_3$}{$n_2$}. \DMethod{SanGuardInMid}{$n_1$}{$n_3$}{$n_2$}\ is true iff there exists a \T{SemChild}
    path between $n_1$, $n_3$, between $n_3$ and $n_2$, with the additional constraint of $n_3$ being a sanitizer or guard. 
\end{enumerate}

The difference between the \T{Vulnerability} relation (which \sa populates) and \T{Witness} relations (which we want to find) is highlighted in {\color{red} red} and {\color{ForestGreen} green}. Notice that while defining the \T{Witness} relation, we simply use the existing relations that define the \T{Vulnerability} relation. Thus, we argue that \sawitness can be implemented on top of \sa by using the intermediate relations that \sa is computing.

\lstMakeShortInline[columns=fixed]@

\subsection{Witness Removal}
\label{subsec:witness-removal}

We obtain \safeprogs and witnesses by applying \sawitness to a snapshot of a codebase. Recall that the witnesses block the flow between a source and a sink and thus help make programs  \textit{safe}. Hence, removing these witnesses will make the programs unsafe. Recall also that the witnesses are either sanitizing functions of the form @sanitize(taintedVar)@ or guards of the form @if checkSafe(taintedVar) {executeSink(taintedVar)}@. 
We implement witness-removal perturbations  that precisely remove the guard-checks and sanitizer-functions. Note that our goal here is to generate unsafe programs and corresponding edits that enable learning repair strategies that insert such witnesses. So, while we generate the unsafe programs by perturbation, they should look structurally similar to natural unsafe programs written by the developers, otherwise the repair strategies learned on this artificially generated data through perturbations would not generalize to code in the wild. 
\lstDeleteShortInline@

\begin{figure}
\scriptsize
\begin{minipage}{0.55\textwidth}

\begin{algorithm}[H]
\begin{algorithmic}[1]
\myFunc{\rmGuard}{$\prog_{safe}$} \label{algo:line:rmGuard}
\State $\prog_{unsafe}$ $\gets$ \textsc{Copy}($\prog_{safe}$)
\State \witness $\gets$ $\prog_{unsafe}$.witness
\State \witnesspar $\gets$ \witness.parent
\State \witnessparpar $\gets$ \witness.parent.parent
\If{\witnesspar.type = \textsc{IfStmt}}
\State parindex $\gets$ \textsc{GetChildIndex}(\witnessparpar, \witnesspar)
\State \witnessparpar.children[parindex] $\gets$ \witnesspar.children[1] 
\EndIf{}
\If{\witnesspar.type = \textsc{BinaryExpr}}
\State parindex $\gets$ \textsc{GetChildIndex}(\witnessparpar, \witnesspar)
\State witindex $\gets$ \textsc{GetChildIndex}(\witnesspar, \witness) 
\State nonwitindex $\gets$ 2 - witindex
\State \witnessparpar.children[parindex] $\gets$ \witnesspar.children[nonwitindex] 
\EndIf{}
\State editprog $\gets$ \witnesspar
\State editloc $\gets$ \witnessparpar
\State \algout{$\{\prog_{unsafe}$, \textsc{Edit}(editprog,editloc), $\prog_{safe}\}$}
\end{algorithmic}
\end{algorithm}
\end{minipage}\hfil%
\begin{minipage}{0.45\textwidth}
\begin{algorithm}[H]
\begin{algorithmic}[1]

\myFunc{\rmSan}{$\prog_{safe}$} \label{algo:line:rmSan}
\State $\prog_{unsafe}$ $\gets$ \textsc{Copy}($\prog_{safe}$)
\State \witness $\gets$ $\prog_{unsafe}$).witness
\State \witnesspar $\gets$ \witness.parent
\If{\witnesspar.type = \textsc{AssignExpr}}
\State parindex $\gets$ \textsc{GetChildIndex}(\witnesspar, \witness)
\State \witnesspar.\textsc{DeleteChild}(parindex)
\EndIf{}
\If{\witnesspar.type = \textsc{Expr}}
\State unsanitized $\gets$ \witness.semparent
\State parindex $\gets$ \textsc{GetChildIndex}(\witnesspar, \witness)
\State \witnesspar.children[parindex] $\gets$ unsanitized 
\EndIf{}
\State editprog $\gets$ \witness
\State editloc $\gets$ \witnesspar
\State \algout{$\{\prog_{unsafe}$, \textsc{Edit}(editprog,editloc), $\prog_{safe}\}$}

\end{algorithmic}
\end{algorithm}
\end{minipage}
\captionof{figure}{Sketch of \rmGuard and \rmSan functions}
\label{fig:remove-functions}
\end{figure}

\lstMakeShortInline[columns=fixed]@

\begin{figure}[h] 
\centering
    \begin{subfigure}[t]{0.49\textwidth}
    \usetikzlibrary{automata,shapes,shapes.geometric,arrows,fit,calc}
    \tikzset{->,elliptic state/.style={draw,ellipse}}
        \begin{tikzpicture}[shorten >=1pt, scale = 0.5, transform shape]
        
            \node[elliptic state, align=center] (A) [] {\{$\dots$\}\\\blockstmt};
            
            \node[elliptic state, align=center, below of = A, xshift=0cm, yshift=-1.4cm] (B) [] {if check(taintVar) exec(taintVar)\\\ifstmt};

            \node[elliptic state, align=center, left of = B, xshift=-2.5cm, yshift=-2.2cm] (C) [] {check(taintVar)\\\callexpr};
            \node[elliptic state, align=center, right of = B, xshift=2.5cm, yshift=-2.2cm] (D) [] {exec(taintVar)\\\callexpr};

            \draw (A) edge[red] node [elliptic state,inner sep=1pt,draw,pos=0.5,left=-3pt,above=0pt, red] {-} (B);
            \draw (B) edge[] (C);
            \draw (B) edge[red] node [elliptic state,inner sep=1pt,draw,pos=0.5,left=-2pt,above=0pt, red] {-} (D);            
            \draw (A) edge[dashed, ForestGreen, bend left=45] node [elliptic state,inner sep=1pt,draw,pos=0.5,left=-2pt,above=0pt, ForestGreen] {+} (D);

        \end{tikzpicture}
        \caption{\rmGuard for guard inside a \ifstmt}
    \label{fig:witness-removal1}
    \end{subfigure}\hfil%
    \begin{subfigure}[t]{0.49\textwidth}
    \usetikzlibrary{automata,shapes,shapes.geometric,arrows,fit,calc}
    \tikzset{->,elliptic state/.style={draw,ellipse}}
        \begin{tikzpicture}[shorten >=1pt, scale = 0.5, transform shape]
        
            
            \node[elliptic state, align=center, align=center] (B) [] {if (taintVar \&\& check(taintVar)) exec(taintVar)\\\ifstmt};

            \node[elliptic state, align=center, left of = B, xshift=-2.5cm, yshift=-2.2cm] (C) [] {taintVar \&\& check(taintVar)\\\binaryexpr};
            \node[elliptic state, align=center, right of = B, xshift=2.5cm, yshift=-2.2cm] (D) [] {exec(taintVar)\\\callexpr};

            \node[elliptic state, align=center, left of = C, xshift=-2.5cm, yshift=-2.2cm] (E) [] {taintVar\\\varexpr};
            \node[elliptic state, align=center, left of = C, xshift=1cm, yshift=-2.2cm] (EF) [] {\&\&\\\binaryop};
            \node[elliptic state, align=center, right of = C, xshift=2.5cm, yshift=-2.2cm] (F) [] {check(taintVar)\\\callexpr};

            \draw (B) edge[red] node [elliptic state,inner sep=1pt,draw,pos=0.5,left=1pt,above=0pt, red] {-} (C);
            \draw (B) edge[] (D);
            \draw (C) edge[red] node [elliptic state,inner sep=1pt,draw,pos=0.5,left=1pt,above=0pt, red] {-} (E);
            \draw (C) edge[] (F);            
            \draw (C) edge[] (EF);            

            \draw (B) edge[dashed, ForestGreen, bend right=40] node [elliptic state,inner sep=1pt,draw,pos=0.5,right=8pt,above=0pt, ForestGreen] {+} (E);

            \end{tikzpicture}
        \caption{\rmGuard for guard inside a \binaryexpr}
    \label{fig:witness-removal2}
    \end{subfigure}\hfil%
    \caption{\rmGuard examples}
    \label{fig:witness-removal}
\end{figure}

We use \rmSan and \rmGuard functions to programmatically remove the witnesses. A high-level sketch of these functions is illustrated in Figure~\ref{fig:remove-functions}. The functions use the structure of the corresponding \astree (node types $\tau$) to decide how to remove witnesses. Consider the \rmGuard function. It first computes the parent (\witnesspar) and grand-parent (\witnessparpar) of the witness guard condition. Then if the type of \witnesspar is \ifstmt (i.e., program is of the form @if (witness) body@ then we modify the \astree edge from \witnessparpar and \witnesspar to instead point to the body of the \ifstmt (index 1 child is body of \ifstmt). Similarly, if the type of \witnesspar is \binaryexpr with operator @&&@ (i.e. of the form @if (otherCond && guard)@ or @if (guard && otherCond)@) then we again modify the edge from \witnessparpar and \witnesspar to instead point to the non-guard child of \binaryexpr (@otherCond@ in the example). Note that since \binaryexpr has 3 children, the index of non-guard child is index of guard-child subtracted from 2. 
Figure~\ref{fig:witness-removal} depicts this removal on the \astree level, where the syntactic edges in red are removed and the syntactic edges in green are inserted.
In the end, the functions returns a tuple of the \pdg of the unsafe program ($\prog_{unsafe}$), \pdg of the safe program ($\prog_{safe}$)
and an edit object (\edit) which stores

\begin{enumerate}
    \item \astree for the removed witness (referred to as \editprog)
    \item location in the \pdg where the witness is removed (referred to as editloc
    or \editloc)
\end{enumerate}

Since $\prog_{unsafe}$ and edit-object can generate the safe program, we only propagate the unsafe programs and edits as the output of this step. Applying \rmGuard function to the safe program in Figure~\ref{fig:safememberex} removes the \ifstmt on Line~\ref{lst:line:fix-start} while preserving the @handlers[callerId](data);@ statement and in fact produces the unsafe program in Figure~\ref{fig:unsafememberex}. Additionally, it  returns the removed witness guard  @if handlers.hasOwnProperty(data.id){ ... }@ as the \editprog and \blockstmt (blue oval in Figure~\ref{fig:example1-pdg}) as the edit location \edit.editloc. Figure~\ref{fig:example1-editprog} shows the \astree for the \editprog containing the \ifstmt. 
The dashed line and dark circle correspond to the \textit{removed} \astree edge between the \blockstmt and the \expr @handlers[callerId](data)@. 

Note that Figure~\ref{fig:remove-functions} provides a high-level sketch of witness-removal and elides over implementation details that are required to make it work for real \js programs. We discuss these issues in the implementation section (Section~\ref{subsec:impl:witness-removal}).
\lstDeleteShortInline@


\section{Strategies and Learning Algorithm}
\label{sec:learning}
In this section, we describe how to learn repair strategies from the  unsafe programs and edits collected in Section~\ref{sec:data}. We define a \dsl (Section~\ref{subsec:dsl}) to express repair strategies that take an \pdg of an unsafe program  as input and generate a safe program as output. The DSL is expressive and can even express bad strategies that don't generalize well to programs in the wild. We provide examples of such bad strategies and good strategies that generalize well  (Section~\ref{subsec:examples}). We learn good repair strategies  in a data-driven manner using an example-based synthesis algorithm (Section~\ref{subsec:synthesis}). 


\subsection{\dsl for repair strategies}
\label{subsec:dsl}
We introduce a novel \dsl to express repair strategies in Figure~\ref{fig:fixing-dsl}.
At a high level, the strategies define a three-step process where  they provide a computation to identify the edit-location node \editloc, a computation to identify the child index $\editindex$ of \editloc where repair happens, and a computation to generate the AST that must be placed at  index $\editindex$ of \editloc for the repair. The main part of these computations involve traversing paths of the input unsafe program \prog.

\setlength{\grammarindent}{5em} 

\begin{figure}
\footnotesize
\begin{align*}
    \DRulenew{@start}{Strategy}{S}\ \text{::=}&\ \DMethodDSL{Insert}{\I{L}}{\I{I}}{\I{O}}\ \ |\ \ \DMethodDSL{Replace}{\I{L}}{\I{I}}{\I{O}} \\
    \DRulenew{}{Node}{L}\ \text{::=}&\  \DMethodDSL{ApplyTraversal}{\I{L}}{\I{F}_k\ o\ \I{F}_{k-1}\ o\ \cdots\I{F}_0}\ \ |\ \ \prog.source\\ 
    \DRulenew{}{GetTraversal}{F}\ \text{::=}&\  \DMethodDSL{GetEdge }{\I{ET}}{\I{I}}\ \ |\ \ \DMethodDSL{GetKleeneStar}{\I{ET}}{\I{C}} \\ 
    \DRulenew{}{GetClauses}{C}\ \text{::=}&\ \DMethodDSL{GetClause}{\tau}\ \ |\ \ \DMethodDSL{GetNeighbourClause}{\I{F}}{\tau}\ \ |\ \ \I{C} \land \I{C} \\
    \DRulenew{} {GetIndex}{I}\ \text{::=}&\ \DMethodDSL{GetConstant}{\I{z}}\ \ |\ \ \DMethodDSL{GetOffsetIndex}{\I{L}, \I{z}} \\
    \DRulenew{}{\eastree}{O}\ \text{::=}&\ \DMethodDSL{ConstantAST}{\tau}{\I{value}}{\I{O}_1}{\I{O}_2}{\cdots}{\I{O}_c}\ \ |\ \ \DMethodDSL{ReferenceAST}{\I{L}} \\
    \DRulenew{}{EdgeType}{ET}\ \text{::=}&\ \texttt{SynParent}\ \ |\ \ \texttt{SynChild}\ \ |\ \ \texttt{SemParent}\ \ |\ \ \texttt{SemChild} \\
    \DRulenew{@input}{Program}{\prog}\ \text{::=}&\ \DMethodDSL{Input}
\end{align*}
\caption{Our \dsl for representing repair strategies. Here, $\tau$ is the set of \astree node types (Section~\ref{sec:data}), \I{value} is the set of possible string representations of \astree.}
\label{fig:fixing-dsl}        
\end{figure}

The top-level production rule of the DSL defines strategies, \strategy, with type \newtextsc{Strategy}. 
\gettraversal, \getclauses, and \getindex are all functions that take a \node $n$ as input and return a \node, \bool, and \integer as output respectively. The edit-AST, \eastree, is similar to a syntactic variant of \astree (i.e. no semantic edges) which we define in Section~\ref{sec:data} with one addition. It has reference nodes that, when applying the strategy to the input \pdg of \prog, are materialized from sub-trees of this \pdg, where the root nodes of these sub-trees are identified by traversing paths in the input. 

The strategy \strategy is of two types, \insertsc or \replace. \DMethod{Insert}{\I{L}}{\I{I}}{\I{O}}\ declaratively expresses the computation that computes the edit-location \editloc by traversing the path supplied in \I{L}, then computes \editindex, the index of edit-location,  by evaluating \I{I}(\editloc), and inserts the materialization of \I{O} as a syntactic-child \astree at index \editindex of the edit-location \editloc. \DMethod{Replace}{\I{L}}{\I{I}}{\I{O}}\ is similar and performs a replacement instead of an insertion.

\node (\I{L}) is either the node corresponding to the source of vulnerability (\prog.source) or the target of the path corresponding to the traversal \DMethod{ApplyTraversal}{L}{\I{F}$_k\ o\ $\I{F}$_{k-1}\ o\ \cdots$\I{F}$_0$}. 
Here, each \I{F}$_i$ 
is a function that takes a node 
$n$ as input, performs a traversal from $n$, and returns the traversal's target node $n'$. 
Thus, \T{ApplyTraversal} can be recursively defined as \DMethod{ApplyTraversal}{\I{F}$_0$(L)}{\I{F}$_k\ o\ $\I{F}$_{k-1}\ o\ \cdots$\I{F}$_1$}\ if $k>0$ and \I{F}$_0$(L) otherwise. 

\newtextsc{GetTraversal} (\I{F}) defines a function that takes a node $n$ and returns a node $n'$ reachable from $n$ and can be of two types. Given $n$, the \DMethod{GetEdge}{\I{ET}}{\I{I}}\ operator first finds the possible single-edge traversals of type \I{ET} and indexes it using \I{I}. Specifically, if edge type \I{ET} is a parent then it returns the parent of $n$. Otherwise, 
it finds a set of $N$ of nodes that are connected with $n$ via the edge type \I{ET}, i.e., $N = \mathcal{E}(n, \I{ET})$, and returns the node $N[I(\I{n})]$ at the index given by $I$. In contrast, $\DMethod{GetKleeneStar}{\I{ET}}{\I{C}}(n)$  performs a \newtextsc{KleeneStarTraversal} that iteratively traverses edges of type \I{ET}, staring from input node $n$, until it reaches an edge whose target  node $n^{i}$ satisfies the condition defined by the clause \I{C}. Formally, \newtextsc{KleeneStarTraversal} can be defined recursively as $KE(n_1,ET,C) = \I{C}(n_1)? n_1 : \left(let\ t\in\mathcal{E}(n_1,ET)\ in\ KE(t,ET,C)\right)$. Here, the node $t$, which is target of an edge with source $n_1$ and type $ET$,  is chosen non-deterministically and our implementation resolves this non-determinism through a breadth-first search.

\newtextsc{GetIndex} (\I{I}) defines a function that takes a node $n$ and returns a \integer. It is either a constant function that returns a fixed integer $z$ or a \DMethod{GetOffsetIndex}{\I{L}, \I{z}}. \DMethod{GetOffsetIndex}{\I{L}, \I{z}}\ takes a node $n$ as input and returns an integer $DO(n,L)+z$, where $DO(n_1,n_2)$ returns the index of syntactic child of $n_2$ who is a syntactic ancestor of $n_1$. 

\eastree (\I{O}) defines the edit \astree with reference nodes which, given an input program \prog, are materialized to a concrete \astree. The \eastree can either be a \T{ConstantAST} or a \T{ReferenceAST}. Specifically, \DMethod{ConstantAST}{$\tau$}{\I{value}}{\I{O}$_1$}{\I{O}$_2$}{$\cdots$}{{\I{O}$_k$}}\ returns an \eastree that has a type $\tau$, string representation \I{value}, and is recursively constructed with sub-trees \I{O}$_1 \cdots$ \I{O}$_k$ as syntactic children, each of which can either be a \T{ConstantAST} or a \T{ReferenceAST}. The \DMethod{ReferenceAST}{\I{L}}, when applying the strategy, finds a node $n$ in \prog by traversing the path described in \I{L} and returns a copy of the (syntactic) sub-tree of \prog rooted at $n$. 

\newcommand{\newwrapbox}[2]{\adjustbox{margin=1pt 1.3pt, bgcolor=white, frame=1pt, cframe=#1, color=#1}{#2}}
\begin{figure}
\centering
\begin{subfigure}{\textwidth}
\begin{mdframed}[,leftmargin=1pt,rightmargin=1pt,innerrightmargin=-3pt,innerleftmargin=7pt,innertopmargin=1pt,innerbottommargin=1pt,roundcorner=7pt]
	\begin{lstlisting}[basicstyle=\scriptsize,numbersep=10pt]
P = input()

Ls = ApplyTraversal(P.source, <@\newwrapbox{newForestGreen}{\textbf{GetKleeneStar}("SemChild", \textbf{GetClause}("VarExpr") $\land$ \textbf{GetNeighbourClause}("Parent", "CallExpr"))}\label{lst:line:semkleene}@>) 
// Ls => foo;
Le = ApplyTraversal(Ls, <@\newwrapbox{newForestGreen}{\textbf{GetKleeneStar}("SynParent", \textbf{GetClause}("BlockStmt"))}\label{lst:line:synkleene}@>)
// Le => { ... } 
I = <@\newwrapbox{newForestGreen}{\textbf{GetOffsetIndex}(Ls, 0)}\label{lst:line:offseteditindex}@>
// I => 13

Lr2 = ApplyTraversal(Ls, <@\newwrapbox{newForestGreen}{\textbf{GetEdge}("SemParent", \textbf{GetConstant}(0)) o \textbf{GetEdge}("SemParent", \textbf{GetConstant}(0))}\label{lst:line:goodref}@>)
// Lr2 => data.id 
Lr1 = ApplyTraversal(Lr2, GetEdge("SynChild", GetConstant(0)) o GetEdge("SynParent", GetConstant(-1)))
// Lr1 => handlers 
Lr3 = ApplyTraversal(Le, GetEdge("SynChild", <@\newwrapbox{newForestGreen}{I}@>))
// Lr3 = foo(data);

O1 = ConstantAST("CallExpr", "...", ConstantAST("DotExpr", "...", ReferenceAST(Lr1), ReferenceAST(Lr2)))
O = ConstantAST("IfStmt", "...", O1, ReferenceAST(Lr3)) <@\label{lst:line:goodstratO}@>
// O => if (handlers.hasOwnProperty(data.id)) { foo(data);}
S = Replace(Le, I, O)

\end{lstlisting}
\end{mdframed}
\caption{Example of a generalizing strategy}
\label{fig:strat1}
\end{subfigure}\hfil%
\begin{subfigure}{\textwidth}
\begin{mdframed}[,leftmargin=1pt,rightmargin=1pt,innerrightmargin=-3pt,innerleftmargin=7pt,innertopmargin=1pt,innerbottommargin=1pt,roundcorner=7pt]
	\begin{lstlisting}[basicstyle=\scriptsize,numbersep=10pt]
P = input()

Ls = ApplyTraversal(P.source, GetEdge("SemChild", GetConstant(0)) o GetEdge("SemChild"  <@\newwrapbox{newred}{$\dots$ 7 times}\label{lst:line:nosemkleene}@> )
// Ls => foo;
Le = ApplyTraversal(Ls, GetEdge("SynParent", GetConstant(-1)) o GetEdge("SynParent"  <@\newwrapbox{newred}{$\dots$ 3 times}\label{lst:line:nosynkleene}@> )
// Le => { ... } 
I = <@\newwrapbox{newred}{\textbf{GetConstant}(13)}\label{lst:line:consteditindex}@>

Lrexpr = ApplyTraversal(Le, <@\newwrapbox{newred}{\textbf{GetEdge}("SynChild", \textbf{GetConstant}(7))}\label{lst:line:badref}@>)
// Lrexpr => let foo = handlers[data.id]
Lr1 = ApplyTraversal(Lrexpr, GetEdge("SynChild", GetConstant(0) o GetEdge("SynChild", GetConstant(1))
// Lr1 => handlers 
Lr2 = ApplyTraversal(Lrexpr, GetEdge("SynChild", GetConstant(1) o GetEdge("SynChild", GetConstant(0))
// Lr2 => data.id 
Lr3 = ApplyTraversal(Le, GetEdge("SynChild", <@\newwrapbox{newred}{I}@>))
// Lr3 = foo(data);

O1 = ConstantAST("CallExpr", "...", ConstantAST("DotExpr", "...", ReferenceAST(Lr1), ReferenceAST(Lr2)))
O = ConstantAST("IfStmt", "...", O1, ReferenceAST(Lr3))<@\label{lst:line:eastree}@>
// O => if (handlers.hasOwnProperty(data.id)) { foo(data);}

S = Replace(Le, I, O)
\end{lstlisting}
\end{mdframed}
\caption{Example of a non-generalizing strategy}
\label{fig:strat2}
\end{subfigure}\hfil%
\caption{Example repair-strategies in our \dsl for the running example in Figure~\ref{fig:vulnerabilty-example1} with key differences highlighted in green and red. We also show the evaluations for the locations corresponding to the example as comments. The strategy on the top generalizes better because it uses semantic edges and \newtextsc{KleeneTraversal}.} 
\label{fig:repair-strategy-ex1}
\end{figure}

\lstMakeShortInline[columns=fixed]@
\subsection{Example of strategies in our \dsl}
\label{subsec:examples}
Figure~\ref{fig:repair-strategy-ex1} describes   two possible repair strategies that are sufficient to repair the motivating example in Figure~\ref{fig:vulnerabilty-example1}. We first describe the good strategy in Figure~\ref{fig:strat1}, referred to as \strategyone,  and then compare it with the bad strategy \strategytwo in Figure~\ref{fig:strat2}. 

Given the program \prog in Figure~\ref{fig:vulnerabilty-example1}(a) as input, the strategy \strategyone
performs a replacement at index \I{I} of edit-location $L_e$ with the materialization of \I{O} (line 20 of \strategyone).
This process requires first finding the "semantic location" node \semloc. 
To this end, the strategy 
first  traverses a path from the node annotated as \T{source} by \sa  using \DMethod{GetKleeneStar}\ in Line~\ref{lst:line:semkleene} of \strategyone.  This \newtextsc{KleeneStarTraversal} starts from \T{source}, traverses semantic dataflow edges, and stops at a node 
corresponding to an identifier being used as the function name in a function call. 
 For the input program $P$, the traversal takes the semantic-child-edges 1-7 (Figure~\ref{fig:example1-pdg}) and stops at @foo@ in Line~\ref{lst:line:callerId-sink} of Figure~\ref{fig:vulnerabilty-example1}(a). Next, to reach the edit-location $L_e$, the strategy uses a \newtextsc{KleeneStarTraversal} that starts from \semloc, traverses syntactic parent edges,  and stops when it reaches a \blockstmt. For $P$, this traversal sets $L_e$  as the node corresponding to the \blockstmt between Lines~\ref{lst:line:handlers-run} and ~\ref{lst:line:handlers-run-end} of Figure~\ref{fig:vulnerabilty-example1}(a). Next, in Line~\ref{lst:line:offseteditindex} of \strategyone, the index \I{I} is set to the index corresponding to the  syntactic child of the edit-location $L_e$ who is an ancestor of the semantic location $L_s$ . For $P$, this index  materializes into $13$; the edge  outgoing from blue \blockstmt in Figure~\ref{fig:example1-pdg} to an ancestor of semantic location (shown in red) has label \T{ch:13}. Next, we materialize the \eastree defined in Line~\ref{lst:line:eastree} of \strategyone by  materializing the  reference-nodes. The \eastree \I{O} serializes into @if (REF1.hasOwnProperty(REF2)) { REF3 } @ where @REF1@, @REF2@, and @REF3@ correspond to \T{ReferenceAST} operators with locations as \reflocone, \refloctwo, and \reflocthree. \refloctwo traverses semantic-parent edges  from \semloc (Line~\ref{lst:line:goodref}) and materialize into @data.id@. Similarly, \reflocone and \reflocthree traverse syntactic children edges and materialize into @handlers@ and @foo(data);@ respectively. Thus, the \eastree \I{O} materializes  into @if (handlers.hasOwnProperty(data.id)) { foo(data); }@, which is the required repair.

Now consider the repair strategy \strategytwo in Figure~\ref{fig:strat2}. This strategy shares a similar structure with the earlier strategy but differs in the way traversals and the index $\I{I}$ are computed. There are four key differences
\begin{enumerate}
    \item In order to reach \semloc from \prog.source, \strategytwo performs the \T{EdgeTraversal} using semantic-child edge seven times in Line~\ref{lst:line:nosemkleene}. The number of semantic edges varies widely across programs and prevents generalization to other scenarios. \T{KleeneStarTraversal} operator instead uses \newtextsc{Clauses} over nodes to find the edit-location.
    \item To reach $L_e$ from \semloc, \strategytwo performs the \T{EdgeTraversal} using syntactic-parent edge seven times in Line~\ref{lst:line:nosynkleene}. Consider a program that instead assigns output of the function-call @let out = foo(data)@. \strategytwo will find \assignexpr as the edit-location and fail to generalize whereas \strategyone will appropriately adjust and take four parent steps.
    \item In order to compute the index at which replacement needs to occur, \strategytwo uses a \DMethod{ConstantIndex}{13}\ in Line~\ref{lst:line:consteditindex} of Figure~\ref{fig:strat2}, which effectively assumes that replacement should always occur at 13$^{th}$ child of $L_e$ and again doesn't generalize. \strategyone on the other hand uses of \T{GetOffsetIndex} operator to instead compute index dynamically for a given input program
    \item In order to materialize reference nodes, \strategytwo uses syntactic edge traversals (Line~\ref{lst:line:badref} of Figure~\ref{fig:strat2}) which assume definite structure about the structure of the program (@GetConstant(7)@ used as syntactic child index to solve a long-ranged-dependency). \strategyone instead uses semantic-parent edges to capture the semantics here and produces a better generalizing repair.
\end{enumerate} 

\lstDeleteShortInline@

\noindent These programs highlight that our \dsl is expressive enough to perform complicated non-local repairs in a generic manner. At the same time, while many strategies can repair a given program, all applicable strategies are not equally good. A key realization is that we \emph{prefer shorter traversal functions} (\newtextsc{KleeneStarTraversal}\ over a long sequence of \newtextsc{EdgeTraversal}). Similarly, we \textit{prefer the traversals with none or small constants}. For example, we prefer \DMethod{GetOffsetIndex}{\semloc}{0}\  over \DMethod{GetConstant}{13}\ and semantic-parent traversal over syntactic-parent traversal with index \DMethod{GetConstant}{7}. 
We use these insights to guide the search in our synthesis algorithm.

\subsection{Synthesizing \dsl strategies from examples}
\label{subsec:synthesis}

Given this high-level \dsl, we will now describe our example-based synthesis algorithm. 
We take as input a set of unsafe programs and edits generated as output at the end of data collection step (Section~\ref{sec:data}). 
Let $\{(\prog_{1},\edit_1),(\prog_{2},\edit_2),\dots$ $,(\prog_{n},\edit_n)\}$. 
Here, $\prog_{i}$ is the $i^{th}$ unsafe program and $\edit_i$ is the corresponding edit. Edit ($\edit$) contains the \astree-node of the edit-location ($\edit$.loc), the \textit{concrete} \astree of the edit-program ($\edit$.editprog), and the type of edit i.e. \insertsc or \replace ($\edit$.type). We use these to learn high-level repair strategies in our \dsl. 
\lstMakeShortInline[columns=fixed]@
Our goal is to combine specific paths, learned over examples that share similar repairs in different semantic and syntactic contexts, to obtain general strategies. Our repair strategies abstractly learn the following:
\begin{enumerate}
    \item Traversals for localizing edit-locations (\editloc) and reference-locations (\refloc). For example, @Ls@ in Line~\ref{lst:line:semkleene} (Strategy \strategyone) depicts a \T{KleeneTraversal} abstraction we can learn from examples having a variable number of semantic-edges. Similarly, @I@ in Line~\ref{lst:line:offseteditindex} (of \strategyone) depicts a generalized index expression we can learn from examples.
    \item \eastree which use reference-traversals. For example, @O@ in Line~\ref{lst:line:goodstratO} demonstrates templated-program-structure that we can learn from examples (say by generalizing from the witnessed guards @handlers.has(data)@  and @events.storage.has(event.name)@).
\end{enumerate}
\lstDeleteShortInline@


We depict our synthesis algorithm in Figure~\ref{fig:strategy-learning}. At a high-level, our synthesis algorithm, first pre-processes the inputs, storing the required \textit{concrete} traversals. Next, it performs ranked pair-wise merging over the processed edits to synthesize strategies.

\noindent \textbf{Pre-processing.} In this step, given the programs and edits, we store the concrete traversals required for learning \editloc and \refloc (Line~\ref{algo:line:preprocess}). Naively computing all such traversals is very expensive and also leads to \textit{bad strategies}. Here, based on the insights from Section~\ref{subsec:examples}, we only compute the traversals that lead to shorter  
traversals
which generalize better. In addition, we also share traversals between between \editloc and \refloc. Pre-processing has following three key steps:
\begin{enumerate}
    \item \textbf{Edit Traversals.} We compute the traversals between \prog.source and \editloc (Line~\ref{algo:line:conceditloc} of Figure~\ref{fig:strategy-learning}) that have the form of a sequence of semantic-edges followed by a sequence of syntactic-edges. This allows abstracting variable-length sequences of semantic-edge traversals as a \kleeneedge (corresponding to an abstract \newtextsc{KleeneTraversal}). We implement this using \newtextsc{BiDirecBFS} method at Line ~\ref{algo:line:bidirecbfs}. For every edit-traversal ($\I{T}_e$), we define {\em semantic-location} (\semloc for brevity) as the last-node on the semantic (dataflow) traversal before traversing the syntactic-edges.
    \item \textbf{Compressing Edit Traversals.} We compress these edit-traversals using the \newtextsc{Compress} method in Line~\ref{algo:line:compress}. It takes in a sequence of (syntactic or semantic) edges as input, greedily combines the consecutive edges with the same edge-type (\edgetype) into a \kleeneedge. The \kleeneedge is constructed using the edge type \edgetype, and a set of clauses $\clause_i$ that satisfy the target node of \kleeneedge. These clauses are either $\lambda n. \mathcal{T}(n) = \tau$ that check the type  or $\lambda n. \mathcal{T}(F_i(n)) = \nu$ that check the type of a neighbor. \newtextsc{Compress} returns a sequence of edges or \kleeneedges as output. 
    \item \textbf{Reference Traversals.} For every node of the edit-program, we locate nodes in the \pdg with the same \textit{value} using a \newtextsc{LevelOrderBFS} until a max-depth (Line~\ref{algo:line:maxlevel}). We perform this traversal from \semloc (defined in (1) above). We thus share parts of traversals between locating \editloc and \refloc which helps in learning \textit{better strategies}. The motivation behind using \semloc is that the expressions necessary for repair will be close to \semloc as it lies on the information-flow path. 
\end{enumerate}

\noindent \textbf{Strategy Synthesis.} Given the edits and the associated traversal meta-data, we synthesize the strategy by pair-wise merging  (Line~\ref{algo:line:callmerge}). \newtextsc{MergeEdits}, the top-level synthesis method, takes a pair of edits as inputs and returns a list of strategies satisfying the example edits. We synthesize the strategies recursively using a deductive search over the non-terminals of the DSL (Figure~\ref{fig:fixing-dsl}). Specifically, to synthesize an expression corresponding to a non-terminal, we deduce which production to use and recursively synthesize the non-terminals given by its production-rule. This has the following key components: 
\begin{enumerate}
    \item \newtextsc{MergeEdits}: It takes pairs of edits as inputs and returns the strategy. It recursively synthesizes the traversal (for \editloc), index, and \eastree. It combines and returns them using the edit-type. 
    \item \newtextsc{MergeTraversal}: It takes two concrete traversals (sequence of edges or \kleeneedges) as inputs and returns the abstracted traversal. by merging elements in the sequence.
    \item \newtextsc{MergeEdge}: It takes two edges or \kleeneedges as inputs and returns a \T{GetKleeneTraversal} or \T{GetEdgeTraversal}. We combine two \kleeneedges using their edge-types and intersecting the clauses stored during pre-processing. We combine two edges using their edge-types, and recursively combining their indices.
    \item \newtextsc{MergeIndex}: It takes two integer indices as inputs and returns an abstracted index. If the two input indices are equal, we return a \T{GetConstant} operator with the input index value. Otherwise, we compute offset as the difference between input-index and index of child of $n$ which has \semloc as descendent (computed by $DO(n, \semloc)$). We return this offset if they are equal and an empty-list otherwise. 
    \item \newtextsc{MergeProg}: It takes two programs as input and returns a list of \eastree, where each list element can materialize into the input programs. If the top-level node in the programs have equal values and types, we combine them as a \T{ConstantAST}. Otherwise, we recursively combine their children. Finally, we merge the reference-traversals corresponding to the input programs and combine them into \T{ReferenceAST}.
\end{enumerate}

Our synthesis procedure is inspired by anti-unification~\cite{anti-unification} and we abstract the paths and edit-programs across different examples. Specifically, our \T{KleeneTraversal} and \T{OffsetIndex} functions allow generalization across paths having different number of edges and different indices where naive abstractions fail. Similarly, \eastree also resemble anti-unification over tree-edits. However, again we use traversals over syntactic and longer-context semantic-edges, for better generalizations  and repairs. 

Finally, note that while we perform pair-wise merges over the edits, the strategy synthesis algorithm can be extended to merge bigger cluster of edits together as well. However, from our experience, we find that the pair-wise merging performs well and is sufficient for our experiments. 
\begin{figure}
\scriptsize
\begin{minipage}{0.49\textwidth}

\begin{algorithm}[H]
\begin{algorithmic}[1]

\myFunc{Learn}{$I \equiv \{(\prog_{1},\concedit_{1}),(\prog_{2},\concedit_{2}),\dots,(\prog_{n},\concedit_{n})\}$} \label{algo:line:learn}
\For{$(\prog,\concedit) \in I$}
    \State \textsc{Preprocess}($\prog,\concedit$)
    
\EndFor
\For{$i, j \in$ \textsc{RankSimilar}($I$)}
    \State \algout{ \textsc{MergeEdits}($\concedit_{i},\concedit_{j}$)} \label{algo:line:callmerge}
\EndFor
\Statex

\myFunc{Preprocess}{$\prog,\concedit$} \label{algo:line:preprocess}
\State traversals $\gets$ \textsc{GetEditLocTraversals}($\prog,\concedit$)
\State $\concedit$.traversals $\gets$ traversals
\State $\concinsertcode\ \gets\ \concedit$.editcode
\For{node $\in \concinsertcode$}
    \State node.refs $\gets$ \emptydict
    \For{$\I{T}_E \in$ traversals}
        \State refs $\gets$ \textsc{MaxLevelBFS}($\I{T}_E$.semLoc, node.value) \label{algo:line:maxlevel}
        \State node.refs[$\I{T}_E$.semLoc] $\gets$ refs
    \EndFor
\EndFor
\Statex

\myFunc{GetEditLocTraversals}{$\prog,\concedit$} \label{algo:line:conceditloc}
\State $\editloc \gets \concedit$.editloc
\State source $\gets$ $\prog$.source
\State traversals $\gets$ \textsc{BiDirecBFS}(source, $\editloc$) \label{algo:line:bidirecbfs}
\For{$\I{T}_E \in$ traversals}
    \State $\I{T}_E$.semLoc $\gets$ \textsc{GetSemanticLoc}($\I{T}_E$)   
    \State $\I{T}_E$ $\gets$ \textsc{Compress}($\I{T}_e$)
\EndFor
\State \algout{traversals}
\Statex

\myFunc{GetSemanticLoc}{$\I{T}_E \equiv \{e_0, e_1, \dots, e_{z-1}\}$} \label{algo:line:semanticLoc}
\For{$e_i \in \I{T}_E$}
    \If{$e_i$.type="SynChild" or $e_i$.type="SynParent"}
    \State \algout{$e_i$.end}
    \EndIf
\EndFor
\State \algout{$e_{z-1}$.end}
\Statex

\myFunc{Compress}{$\I{T}_E \equiv \{e_0, e_1, \dots, e_{z-1}\}$} \label{algo:line:compress}
\State ret $\gets$ \emptylist
\State $i \gets 0$
\While{$i < z $}
    \If{$e_i$.type = SynChild}
        \State ret.add($e_i$)
        \State $i \gets i+1$
        \State \textbf{continue}
    \EndIf
    \For{$j \in \{i,i+1,\dots,z-1\}$}
        \State \textbf{if} $e_j$.type != $e_i$.type : \textbf{break}
    \EndFor
    \If{$j \leq i+1$}
        \State ret.add($e_i$) 
        \State $i \gets i+1$
    \Else
        \If{$j = z-1$}
            \State \I{C} $\gets$ \textsc{GetClauses}($e_j$.end)
        \Else
            \State \I{C} $\gets$ \textsc{GetClauses}($e_j$.start)
        \EndIf
        \State ret.add(\textsc{KleeneEdge}($e_j$.type, \I{C}))
        \State $i \gets j+1$ 
    \EndIf
\EndWhile
\State \algout{ret}
\Statex



\end{algorithmic}
\end{algorithm}

\end{minipage}\hfil%
\begin{minipage}{0.51\textwidth}

\begin{algorithm}[H]
\begin{algorithmic}[1]

\setcounter{ALG@line}{43}

\myFunc{MergeEdit}{$\concedit_{i}, \concedit_{j}$} \label{algo:line:mergeedit}
\State ret $\gets$ \emptylist
\State type $\gets$ $\concedit_{i}$.type \algorithmiccomment{\textsc{Insert}/\textsc{Replace}}
\State traversals $\gets$ \textsc{MergeTraversals}($\concedit_{i}$.traversals, $\concedit_{j}$.traversals)
\For{$T_E \in$ edittraversals}
    \State \textbf{global} semLocs $\gets T_E$.semLocs 
    \State \textbf{global} $T_S \gets T_E$.semTraversal 
    \State editProgs $\gets$ \textsc{MergeProg}($\concedit_{i}$.editprog, $\concedit_{j}$.editprog)
    \State index $\gets$ \textsc{MergeIndex}($\concedit_{j}$.index,$\concedit_{j}$.index, $T_E$.endNodes)
    \For{editProg $\in$ editProgs}
        \State ret.add(\textsc{type}($T_E$, index, editProg)) \label{algo:line:assembleedit}
    \EndFor
\EndFor
\State \algout{ret}
\Statex

\myFunc{MergeTraversals}{$\{$\I{T}$_1^1$,\I{T}$_2^1$,\dots,\I{T}$_n^1\}$, $\{$\I{T}$_1^2$,\I{T}$_2^2$,\dots,\I{T}$_n^2\}$}
\State ret $\gets$ \emptylist
\For{every \I{T}$_i^1$ and \I{T}$_j^2$}
    \State ret.add(\textsc{MergeTraversal}(\I{T}$_i^1$,\I{T}$_j^2$))
\EndFor
\State \algout{ret}
\Statex

\myFunc{MergeTraversal}{$\I{T}_i \equiv \{e_0^i, \dots, e_{n}^i\}$, $\I{T}_i \equiv \{e_0^j, \dots, e_{m}^j\}$}
\State \textbf{if} $n \neq m$ : \algout{\emptylist}
\State mergeEdges $\gets$ \emptydict
\For{$k \in \{1,2,\dots,n\}$}
    \State mergeEdges[k] $\gets$ \textsc{MergeEdge}($e_k^i, e_k^j$)
\EndFor
\State \algout{\textsc{CartesianProduct}(mergeEdges)}
\Statex

\myFunc{MergeEdge}{$e_i$, $e_j$}
\If{$e_i$ and $e_j$ are \textsc{KleeneEdge} and $e_i$.type = $e_j$.type}
    \State \algout{\texttt{GetKleeneTraversal}($e_i$.type, $e_i$.\I{C}$\ \cap\ e_j$.\I{C})}
\EndIf
\If{$e_i$ and $e_j$ are \textsc{Edge} and $e_i$.type = $e_j$.type}
    \State index $\gets$ \textsc{MergeIndex}($e_i$.index, $e_j$.index, [$e_i$.start, $e_j$.start])
    \State \algout{\texttt{GetEdgeTraversal}($e_i$.type, index)}
\EndIf
\Statex

\myFunc{MergeIndex}{$I_i$, $I_j$, [$n_i$, $n_j$]}
\If{$I_i = I_j$}
    \State \algout{\texttt{GetConstant}($I_i$)}
\EndIf
\State offset$_i = I_i - DO(n_i,$ semLocs$_i)$
\State offset$_j = I_j - DO(n_j,$ semLocs$_j)$
\If{offset$_i = $offset$_j$}
   \State \algout{\texttt{GetOffsetIndex}($T_S$, offset$_i$)}
\EndIf
\State \algout{\emptylist}
\Statex

\myFunc{MergeProg}{$\concinsertcode_i,\concinsertcode_j$}
\State ret $\gets$ \emptylist
\If{$\concinsertcode_i$.value = $\concinsertcode_j$.value and $\concinsertcode_i$.type = $\concinsertcode_j$.type}
    \State ret.add({\texttt ConstantAST}($\concinsertcode_i$))
\EndIf
\If{$\concinsertcode_i$.type = $\concinsertcode_j$.type}
\State mergedChildren $\gets$ \textsc{MergeProg}($\concinsertcode_i$.children, $\concinsertcode_j$.children)
\State ret.add({\texttt ConstantAST}($\concinsertcode_i$.type, mergedChildren))
\EndIf
\State refs$_i$, refs$_j$ $\gets$ $\concinsertcode_i$.refs[semLoc$_i$], $\concinsertcode_j$.refs[semLoc$_j$]
\State reftraversals $\gets$ \textsc{MergeTraversal}(refs$_i$, refs$_j$)
\For{$\I{T}_R$ $\in$ reftraversals}
    \State ret.add({\texttt Reference}($\I{T}_R$))
\EndFor
\State \algout{ret}
\Statex



\end{algorithmic}
\end{algorithm}
\end{minipage}
\captionof{figure}{Sketch of our strategy learning algorithm}
\label{fig:strategy-learning}
\vspace{-5pt}
\end{figure}

\section{Implementation}
\label{sec:impl}
\label{subsec:impl:witness-removal}
\lstMakeShortInline[columns=fixed]@

We use \codeql~\cite{codeql} as our \sa tool. It is an open-source tool where custom static analysis is implemented as queries in a high-level object-oriented extension of datalog. The queries follow a relational \Verb|select from where| syntax to query the program database. Thus, we are able to implement the \T{\textbf{Witness}} relation defined in Section~\ref{sec:data} as queries in \codeql.

We implement the witness-removal and strategy learning steps in \cpp. Specifically, we perturb the detected \safeprogs using the \astree structure of the programs as described in Figure~\ref{fig:witness-removal}. While implementing witness-removal, we need to handle two particular challenges
\begin{enumerate}
    \item \textbf{Ensuring naturalness of perturbed programs.} Consider the following program @{if (witness) {sink}}@. Here, during witness-removal, apart from removing the guard condition, we need to remove the additional braces around the @sink@ as well. This is because the corresponding perturbed unsafe program generated (``@{{sink}}@'') would look unnatural and lead to non-generalizing repair strategies. We take care of such corner cases to the best of our abilities and leave investigating a more-thorough witness-removal pipeline for future work.
    \item \textbf{Capturing generalizing edits.} Consider the program @if (!witness) return custommessage@. Here, witness-removal step removes the entire \ifstmt (including the return statement). However, while capturing the edit, we ignore the return-value and store the edit-program as only @if (!witness) return@. This is because the return values, error handling, and error messages are very customized across different codebases and not learnable using a programmatic strategy. We make such design decisions to capture these kinds of \textit{generalizing edits} and discuss the implications in Section~\ref{sec:discuss}.
\end{enumerate}
\lstDeleteShortInline@





\section{Experiments}
\label{sec:experiments}
We present an empirical study of the proposed \sysname\ approach for \js vulnerabilities with \codeql as the \sa\ tool. In particular, we look at two vulnerabilities, \unvalidatedmembership and \xss,  prevalent in \js repositories. The goal of our study is to investigate how well the strategies learned by \sysname\ generalize to code in the wild, and how our method fares in comparison with state-of-the-art techniques for repair.




\subsection{Datasets}
We work with two types of datasets --- one dataset exclusively for training and the other exclusively for evaluation.\footnote{Our datasets can be found at \url{http://aka.ms/StaticFixer
}} 

For \textit{training} (e.g., for learning strategies in \sysname{}), we use \js\ programs in the repositories available on \textsc{LGTM}~\cite{lgtm} that are witnessing-safe (discussed in Section~\ref{sec:data}). We form a dataset (``\benchsmall'') of (safe, unsafe) programs, using our \sawitness technique and \codeql\ queries on \textsc{LGTM}~\cite{lgtm}, for the two vulnerability classes \unvalidatedcall and \xss. This dataset contains \unsure{800} paired programs for \unvalidatedcall and \unsure{66} paired programs for \xss; each pair consists of (i) a \js \safeprog that has a \textit{witness} relation (a sanitizer or a guard, with the corresponding source and sink nodes) discussed in Section~\ref{subsec:sa-witness}, and (ii) the corresponding ``unsafe'' program that is obtained by removing the witness using techniques discussed in Section~\ref{subsec:witness-removal}. 

For \textit{evaluation}, we consider \js code in the repositories on LGTM~\cite{lgtm} (``\benchwild'') that are flagged as vulnerable (\xss or \unvalidatedmembership) by \codeql. We purge all the duplicate files --- e.g., common \js\ libraries are part of multiple repositories. After de-duplication, we have 330 unsafe \js\ files from 204 repositories for \unvalidatedmembership, and 672 unsafe \js\ files from 595 repositories for \xss.


\subsection{Compared Techniques}
The datasets we use are real world \js files, and there is no prior work on repairing 
information flow vulnerabilities in \js\ --- so we cannot use prior work in static repair as baselines (we qualitatively compare against them in Section~\ref{sec:related}). For example, if we were to use \othersysname{Phoenix}~\cite{bavishi2019phoenix}, the most closely related system to \sysname, as a baseline then we would need to (i) port its front-end to consume \js programs instead of Java, and (ii) generalize it to handle the repairs we seek from the more localized repairs it performs, which is a huge engineering effort. Furthermore, \othersysname{Phoenix} implementation is proprietary and not available for this purpose.
 However, fine-tuning {\em neural techniques}  to repair information flow vulnerabilities in \js is feasible and we use them as baselines. The engineering effort required here is tenable as the models are available for public use, there are no front-end issues as they consume program strings, and we only need to provide natural language descriptions and program examples. 

Therefore, we compare \sysname{} with the following state-of-the-art neural techniques for code synthesis, adapted for repair:
\begin{enumerate}
    \item \codetfivejs\ --- we fine-tune the state-of-the-art code synthesis model \codetfive~\cite{Wang2021CodeT5IU} on the training set (\benchsmall), to synthesize fixed code given vulnerable code as input.
    \item \codex\ --- we use few-shot learning on the OpenAI's \codex\ model~\cite{Codex}, to synthesize fixed code given vulnerable code and a set of paired programs (from \benchsmall) as input.
\end{enumerate}

As discussed in Section~\ref{sec:learning}, \sysname solves the problems of both localization and repair. However, given a source-code file, the neural baselines do not have the ability to localize the program statements which need to be transformed for repairing the vulnerabilities. So, we parse the CodeQL warnings and provide the functions identified by CodeQL in its results as part of the input to the neural models. We give more details below. \\
\textbf{Implementation details:} We implement \codetfivejs\ as follows. We use the {\tt CodeT5-small} variant of the model made up of 60M parameters initialised with pre-trained weights. 
For each of the two vulnerability classes, we finetune the model on the corresponding \benchsmall dataset to produce fixed code given vulnerable code.
Given a vulnerable file, we use \codeql to identify function where the sink is, and pass this function as input to the model.
We then finetune the model to produce non-vulnerable code (i.e., the corresponding sink function from the safe program in the training data), given the (localized) vulnerable code.
During evaluation, we use beam-search decoding with a beam size of 20 and a temperature of 1.0, and generate 20 candidate snippets for each input. This ensures a fair comparison to \sysname{} where multiple strategies are applied to a given vulnerable code to generate multiple candidates (for 95\% of the files in~\benchwild,~\sysname{} generates at most 17 candidate fixes). 


For the \codex\ model, we use a subset of unsafe, safe program pairs chosen from the \benchsmall dataset to assemble a ``prompt'' encoding a description of the task, followed by the actual ``question'' (i.e., the vulnerable code) to generate output. 
In the experiments, we construct a prompt of the form $\{d^v, (u^v_1, s^v_1), \dots, (u^v_k, s^v_k), u^v_q\}$; where $d^v$ is a natural language description of how the vulnerability $v$ should be fixed (taken from the CodeQL documentation), $u^v_i$ and $s^v_i$ are unsafe (vulnerable) and the corresponding safe code snippets that demonstrate how the vulnerability should be fixed, and finally $u^v_q$ which is the \textit{localised} vulnerable code snippet that we want the model to fix.
The vulnerability-fixing $(u^v_i, s^v_i)$ examples are drawn from a list of manually-chosen fixes (sink functions, to be consistent with \codetfivejs\ model) from \benchsmall; \codex~has a limit of 8000 input tokens, so the number of such examples in the prompt is determined dynamically depending on the size of $u^v_q$.
We use a temperature setting of 0.9 and generate top 20 candidate outputs ranked by their probability. 








\subsection{Metrics}
We report the number of successful fixes --- we count an unsafe (as flagged by \codeql) code as fixed if the corresponding output code from a method passes the vulnerability check of \codeql. Additionally, \unsure{we manually inspect the generated code to check if there are any unintended changes introduced in the original code. We deem the candidate fixes unsuccessful if there are any unintended changes or if they have any syntactic errors.} 

\subsection{Results}
\label{sec:benchwild}
The number of successful fixes obtained via our method and the baselines on the \benchwild~dataset is reported in Table~\ref{tab:resultswild}. Recall that, for all the methods, we use the unsafe, safe program pairs from the \benchsmall{} dataset for training. It is clear from the results that~\sysname{} is not only significantly better than the baselines relatively but is also highly accurate in an absolute scale. In particular,~\sysname{} (a) generates a successful fix (out of the possibly multiple fixes generated) for nearly 94\% of the (vulnerable) files in the \unvalidatedmembership~class and for nearly 92\% of the files in the \xss~class, and (b) significantly outperforms, by as much as 2.5x, both the state-of-the-art neural techniques in both the vulnerability classes.

Even though we provide additional context to help the neural models localize (such as functions in the CodeQL warnings), these models fundamentally do not try to encode or exploit the domain knowledge. Fine-tuning very large neural models typically needs thousands, if not more, of training examples to be able to generalize well. However, in the real world, it is extremely challenging to collect training data of such scale without significant human effort. 

Consider the following function in \benchwild that is flagged for \unvalidatedmembership vulnerability in line 2:

\begin{lstlisting}[autogobble,basicstyle=\scriptsize,xleftmargin=.24\textwidth, xrightmargin=.2\textwidth]
router.get('/api/crawlers/:type/:username', async (ctx) => {
  const ojFunc = crawlers[ctx.params.type]
  if (!_.isFunction(ojFunc)) {
    throw new Error('Crawler of the oj does not exist')
  }
  ctx.rest(await ojFunc(ctx.params.username))
})
\end{lstlisting}
\lstMakeShortInline[columns=fixed]@

When passed as input to \codetfivejs, we observe that the top candidate fixes it generates fall into two categories: (a) adding a redundant type check after line 5 such as @if (typeof ojFunc === 'function') { ojFunc = crawlers[ctx.params.type] }@, or (b) adding an incorrect membership check such as @if (ctx.params.username in crawlers)@ at line 2. Even though (b) captures the structure of the desired fix, it is not semantically correct, i.e., it checks for the wrong member @username@ instead of @type@. We find that many of the unsuccessful cases for the neural models have similar failure modes --- it reflects the inability of the neural models to capture both the structural and semantic contexts needed to produce intended repairs, for real-world scenarios. \sysname{}, by design, takes into account the data flow information and the semantics and produces semantically correct fixes in a vast majority of cases. In particular, for this example, it generates @if (crawlers.hasOwnProperty(ctx.params.type)) {ojFunc = crawlers[ctx.params.type];}@, at line 2.
\lstDeleteShortInline@
We also notice from Table~\ref{tab:resultswild} that \codex performs poorly on~\xss vulnerability with the most common failure case being that it copies over the vulnerable input as the output.




The success of our method can be attributed to two factors: a) our DSL provides a rich space of strategies, b) our learning algorithm learns a diverse set of strategies (\unsure{292} in total for \unvalidatedmembership, and \unsure{28} for \xss) guided by the limited set of ``perturbed'' safe programs and the witnesses in the training set. This diversity of strategies helps generalize to programs in the wild, which may deviate significantly in size, structure, and semantics from those in the training dataset. Furthermore, we find that, on average,~\sysname{} produces about 4 unique fixes for a given vulnerable code in the~\benchwild dataset, of which about 3 are successful. Thus,~\sysname{} is not only successful on a vast majority of the test files, but also produces multiple, unique correct fixes. This can be especially helpful in practice, when factors besides correctness can determine the suitability of a fix. 


\begin{table*}[!t]
    \centering
    \begin{tabular}{l|c|c}
       \toprule
       Method & \unvalidatedmembership & \xss  \\ \toprule
       \codetfivejs &  127 (38.48\%)& 541 (80.51\%)\\
       \codex &  220 (66.67\%)& 219 (32.59\%)\\
       \sysname &  \textbf{310} (93.94\%)& \textbf{617} (91.82\%)\\ \toprule
    \end{tabular}
    \caption{Number of successful fixes by various methods on the \benchwild\ dataset, out of (i) 330 \js files for \unvalidatedmembership, and (ii) 672 \js files for \xss. All the methods are trained on the~\benchsmall dataset.}
    \label{tab:resultswild}
    \vspace{-25pt}
\end{table*}

\section{Related Work}
\label{sec:related}
Automatic program repair is an active area of research. 
We refer the reader to~\cite{survey1, survey2} for broad survey. Below, we list and compare related works that use static analysis as well as other approaches.\\
\textbf{Static Analysis Based Repair.} 
Systems such as \othersysname{FootPatch}~\cite{FootPatch} and \othersysname{SenX}~\cite{senx} utilize static-analysis-information to create repair strategies. Specifically, \othersysname{FootPatch} reasons about semantic properties of programs and \othersysname{SenX} determines safety properties being violated to generate the patches. 
Unlike these approaches, \sysname  uses static-analysis information (namely witnesses) to generate a paired unsafe-safe dataset of programs and then uses program synthesis to learn repair strategies from the dataset.
Prior work uses static analysis to detect bugs, and then use cross-commit-data from manual fixes to create paired unsafe-safe dataset of programs, and learn repair strategies.
For instance,
\othersysname{SpongeBugs}~\cite{sonarcube} uses \othersysname{SonarCube}~\cite{sonarsa} static analysis to find bugs, and cross-commit data to create paired dataset.
Similarly,~\othersysname{Avatar}~\cite{Liu:mining,liu2019avatar} uses \othersysname{FindBugs}~\cite{findbugs} static analysis and cross-commit data.
\othersysname{GetAFix}~\cite{bader2019getafix} similarly mines general tree-edit-patterns from cross-commit data using anti-unification.
However, these fix-templates or edit patterns are purely syntactic whereas our repair strategies use semantic knowledge 
 (specifically data flow) of programs, which is necessary for fixing information flow vulnerabilities. 
 The \othersysname{Phoenix} tool~\cite{bavishi2019phoenix} makes more use of semantic information, and is closest to our approach in terms of learning repair strategies. However, we use \sawitness to learn repair strategies from a single snapshot of codebases whereas all previous approaches including \othersysname{Phoenix} run \sa across all historical commits of the repository to get paired cross-commit data, which is inherently noisy. Getting clean paired data from commits for bug fixes is a difficult problem, and we avoid this problem entirely. Additionally, our \dsl supports \kleeneedge based operators that allow learning general repairs across examples where data flow paths have variable lengths, which is  not supported by \othersysname{Phoenix}.  

\noindent \textbf{Other Automated Program Repair approaches.} 
\othersysname{Blade}~\cite{blade} and \othersysname{Lifty}~\cite{lifty} repair information-leaks in programs using type analysis. \othersysname{HyperGI}~\cite{hypergi} performs repair on information-leak bugs using test suites.
\othersysname{Refazer}~\cite{rolim2017learning,zhang2022overwatch} learns program transformations from developer edits using program synthesis. \othersysname{VurLe}~\cite{vurle} and \othersysname{Seader}~\cite{exampleBasedJava} both learn program repairs from examples. \othersysname{CDRep}~\cite{cdrep} proposes an approach to repair speculative leaks from cryptographic code. However, the approach requires users to manually write repair templates for cryptographic APIs. \othersysname{BovInspector}~\cite{bovinspector} implements guard templates for fixing buffer overflow in \othersysname{C} programs. Automated program repair techniques have used mutations of buggy programs to pass test cases in a suite ~\cite{GenProg:ICSE2012, ACS:ICSE2017, SearchRepair:ASE2015, ssFix:ASE2017, genesis, Prophet}. More recently, machine learning-based techniques have also started to gain attention for performing repair~\cite{graphbased, allamanis2021self, yasunaga2021break}. Our neural baselines emulate the recent advancements in neural large language models for code-generation, repair, etc. ~\cite{LLMrepair, Xia2022PracticalPR}.



\section{Discussion}
\label{sec:discuss}
\begin{figure}
\centering
\begin{subfigure}{0.48\textwidth}
	\begin{lstlisting}[basicstyle=\scriptsize]
app.get('/perform/action', function(req, res) {
  let action = actions[req.params.action];
  <@{\LARGE \dots}@>
<@\INSp{\ if actions.hasOwnProperty(req.params.action)\{}@>
<@\mINS{\ \ \ \ \ \ \ \ res.end(action(req.params.payload));}@>
<@\INSp{\ \}}@>
});
\end{lstlisting}
\caption{Example for fixed safe program}
\label{fig:imperfect-fix1}
\end{subfigure}
\begin{subfigure}{0.48\textwidth}
	\begin{lstlisting}[basicstyle=\scriptsize]
app.get('/perform/action', function(req, res) {
  let action = actions[req.params.action];
<@\INSp{\ if (!actions.hasOwnProperty(req.params.action))\{}@>
<@\INSp{\ \ \ \ \ \ return;}@>
<@\INSp{\ \}}@>
  res.end(action(req.params.payload));
});
\end{lstlisting}
\caption{Example for fixed safe program}
\label{fig:imperfect-fix2}
\end{subfigure}
\vspace{-5pt}
\caption{Examples of fixes where broader application context is required to predict \textit{natural} fixes}
\label{fig:imperfect-fixes}
\vspace{-18pt}
\end{figure}


    
\lstMakeShortInline[columns=fixed]@

We propose a novel approach to use static analysis and a repository of correct programs that satisfy a property, to automatically learn  strategies to repair programs that violate the property.
We have implemented our approach in the \sysname system.
We evaluate our approach by performing repairs on two specific \js vulnerabilities (unvalidated dynamic call and cross-site scripting) and learn general repair strategies. These repair strategies are able to automatically repair over 90\% of the violations of these properties we found in over 1000 files collected from open-source repositories.


\out{
Our work makes two distinct contributions: (1) to use static analysis witnesses to learn paired datasets, (2) new program synthesis techniques (including novel domain specific language, strategy learning methods, etc) to learn non-local strategies from these datasets. We can potentially use the datasets collected from our approach to train or fine-tune better repair strategies using neural networks~\cite{Codex,Austin}, even though the synthesis approach outperforms neural baselines (see Section 6). 
}
Our approach has two known limitations that can be potentially addressed in future work. 
The first limitation is due to our current implementation architecture.
While our \astree implementation can trace data flows across method boundaries, our \astree is limited to a single file.
A better \astree builder would allow \sysname{} to repair flow vulnerabilities that cross file boundaries.
The second limitation is a conceptual one.
In addition to introducing sanitizers and guards judiciously, repairing information flow violations may also require application-specific side-effect handling, which is beyond the scope of this paper. For example, in Figure~\ref{fig:imperfect-fix1}, the guard blocks the dynamic execution of the function call to avoid the vulnerability. However, real-world fixes would also require appropriate error handling for the "else" branch, such as sending a suitable error message. The repair shown in Figure~\ref{fig:imperfect-fix2} suffers from the same issue, where it terminates function execution via a @return@ without any error message or returning an error value. We imagine a human-in-the-loop repair process where our \sysname suggests the repair witnesses and human reviewers judge the repairs and additionally handle context-specific side effects such as error handling. We also envision a neuro-symbolic program repair system where the broader application context is \textit{predicted} by a neural model like Codex~\cite{Codex} as a future direction.

Though we have evaluated our approach on two specific instances information flow properties, our approach has the potential to repair many classes of information-flow vulnerabilities such as null-dereferencing~\cite{nullowasp}, zip-slips~\cite{zipslipcodeql}, tainted-path~\cite{taintedpathcodeql}, SQL and Code Injection.

\bibliography{main}


\begin{thebibliography}{45}


\ifx \showCODEN    \undefined \def \showCODEN     #1{\unskip}     \fi
\ifx \showDOI      \undefined \def \showDOI       #1{#1}\fi
\ifx \showISBNx    \undefined \def \showISBNx     #1{\unskip}     \fi
\ifx \showISBNxiii \undefined \def \showISBNxiii  #1{\unskip}     \fi
\ifx \showISSN     \undefined \def \showISSN      #1{\unskip}     \fi
\ifx \showLCCN     \undefined \def \showLCCN      #1{\unskip}     \fi
\ifx \shownote     \undefined \def \shownote      #1{#1}          \fi
\ifx \showarticletitle \undefined \def \showarticletitle #1{#1}   \fi
\ifx \showURL      \undefined \def \showURL       {\relax}        \fi
\providecommand\bibfield[2]{#2}
\providecommand\bibinfo[2]{#2}
\providecommand\natexlab[1]{#1}
\providecommand\showeprint[2][]{arXiv:#2}

\bibitem[zip({[n.\,d.]})]%
        {zipslipcodeql}
 \bibinfo{year}{[n.\,d.]}\natexlab{}.
\newblock \bibinfo{title}{Arbitrary file write during zip extraction (Zip
  Slip)}.
\newblock
  \bibinfo{howpublished}{\url{https://codeql.github.com/codeql-query-help/javascript/js-zipslip/}}.
\newblock
\newblock
\shownote{Accessed: Nov 10, 2022}.


\bibitem[cod({[n.\,d.]})]%
        {codeqlast}
 \bibinfo{year}{[n.\,d.]}\natexlab{}.
\newblock \bibinfo{title}{CodeQL AST Types}.
\newblock
  \bibinfo{howpublished}{\url{https://codeql.github.com/docs/codeql-language-guides/abstract-syntax-tree-classes-for-working-with-javascript-and-typescript-programs/}}.
\newblock
\newblock
\shownote{Accessed: Nov 10, 2022}.


\bibitem[was({[n.\,d.]})]%
        {wasp}
 \bibinfo{year}{[n.\,d.]}\natexlab{}.
\newblock \bibinfo{title}{Cross-Site-Scripting}.
\newblock
  \bibinfo{howpublished}{\url{https://owasp.org/www-community/attacks/xss/}}.
\newblock
\newblock
\shownote{Accessed: Nov 10, 2022}.


\bibitem[fin({[n.\,d.]})]%
        {findbugs}
 \bibinfo{year}{[n.\,d.]}\natexlab{}.
\newblock \bibinfo{title}{FindBugs Project}.
\newblock \bibinfo{howpublished}{\url{https://spotbugs.github.io/}}.
\newblock
\newblock
\shownote{Accessed: Nov 10, 2022}.


\bibitem[lgt({[n.\,d.]})]%
        {lgtm}
 \bibinfo{year}{[n.\,d.]}\natexlab{}.
\newblock \bibinfo{title}{{LGTM}}.
\newblock \bibinfo{howpublished}{\url{lgtm.com}}.
\newblock
\newblock
\shownote{Accessed: Oct 1, 2022}.


\bibitem[nul({[n.\,d.]})]%
        {nullowasp}
 \bibinfo{year}{[n.\,d.]}\natexlab{}.
\newblock \bibinfo{title}{Null Dereferencing}.
\newblock
  \bibinfo{howpublished}{\url{https://owasp.org/www-community/vulnerabilities/Null_Dereference}}.
\newblock
\newblock
\shownote{Accessed: Nov 10, 2022}.


\bibitem[xss({[n.\,d.]})]%
        {xsscodeql}
 \bibinfo{year}{[n.\,d.]}\natexlab{}.
\newblock \bibinfo{title}{Reflected cross-site scripting}.
\newblock
  \bibinfo{howpublished}{\url{https://codeql.github.com/codeql-query-help/javascript/js-reflected-xss/}}.
\newblock
\newblock
\shownote{Accessed: Nov 10, 2022}.


\bibitem[son({[n.\,d.]})]%
        {sonarsa}
 \bibinfo{year}{[n.\,d.]}\natexlab{}.
\newblock \bibinfo{title}{SonarQube}.
\newblock \bibinfo{howpublished}{\url{https://docs.sonarqube.org/latest/}}.
\newblock
\newblock
\shownote{Accessed: Nov 10, 2022}.


\bibitem[sql({[n.\,d.]})]%
        {sqlinjowasp}
 \bibinfo{year}{[n.\,d.]}\natexlab{}.
\newblock \bibinfo{title}{SQL Injection}.
\newblock
  \bibinfo{howpublished}{\url{https://owasp.org/www-community/attacks/SQL_Injection}}.
\newblock
\newblock
\shownote{Accessed: Nov 10, 2022}.


\bibitem[tai({[n.\,d.]})]%
        {taintedpathcodeql}
 \bibinfo{year}{[n.\,d.]}\natexlab{}.
\newblock \bibinfo{title}{Uncontrolled data used in path expression (Tainted
  Path)}.
\newblock
  \bibinfo{howpublished}{\url{https://codeql.github.com/codeql-query-help/javascript/js-path-injection/}}.
\newblock
\newblock
\shownote{Accessed: Nov 10, 2022}.


\bibitem[Allamanis et~al\mbox{.}(2021)]%
        {allamanis2021self}
\bibfield{author}{\bibinfo{person}{Miltiadis Allamanis}, \bibinfo{person}{Henry
  Jackson-Flux}, {and} \bibinfo{person}{Marc Brockschmidt}.}
  \bibinfo{year}{2021}\natexlab{}.
\newblock \showarticletitle{Self-Supervised Bug Detection and Repair}. In
  \bibinfo{booktitle}{\emph{NeurIPS}}.
\newblock


\bibitem[Arteau(2018)]%
        {prototypepoll}
\bibfield{author}{\bibinfo{person}{Olivier Arteau}.}
  \bibinfo{year}{2018}\natexlab{}.
\newblock \showarticletitle{Prototype Pollution Attack in NodeJS Application}.
\newblock \bibinfo{journal}{\emph{North Sec}} (\bibinfo{year}{2018}).
\newblock


\bibitem[Avgustinov et~al\mbox{.}(2016)]%
        {codeql}
\bibfield{author}{\bibinfo{person}{Pavel Avgustinov}, \bibinfo{person}{Oege de
  Moor}, \bibinfo{person}{Michael~Peyton Jones}, {and} \bibinfo{person}{Max
  Sch{\"a}fer}.} \bibinfo{year}{2016}\natexlab{}.
\newblock \showarticletitle{{QL: Object-oriented Queries on Relational Data}}.
  In \bibinfo{booktitle}{\emph{30th European Conference on Object-Oriented
  Programming (ECOOP 2016)}} \emph{(\bibinfo{series}{Leibniz International
  Proceedings in Informatics (LIPIcs)}, Vol.~\bibinfo{volume}{56})},
  \bibfield{editor}{\bibinfo{person}{Shriram Krishnamurthi} {and}
  \bibinfo{person}{Benjamin~S. Lerner}} (Eds.). \bibinfo{publisher}{Schloss
  Dagstuhl--Leibniz-Zentrum fuer Informatik}, \bibinfo{address}{Dagstuhl,
  Germany}, \bibinfo{pages}{2:1--2:25}.
\newblock
\showISBNx{978-3-95977-014-9}
\showISSN{1868-8969}
\urldef\tempurl%
\url{https://doi.org/10.4230/LIPIcs.ECOOP.2016.2}
\showDOI{\tempurl}


\bibitem[Bader et~al\mbox{.}(2019)]%
        {bader2019getafix}
\bibfield{author}{\bibinfo{person}{Johannes Bader}, \bibinfo{person}{Andrew
  Scott}, \bibinfo{person}{Michael Pradel}, {and} \bibinfo{person}{Satish
  Chandra}.} \bibinfo{year}{2019}\natexlab{}.
\newblock \showarticletitle{Getafix: Learning to Fix Bugs Automatically}.
\newblock \bibinfo{journal}{\emph{Proc. ACM Program. Lang.}}
  \bibinfo{volume}{3}, \bibinfo{number}{OOPSLA}, Article
  \bibinfo{articleno}{159} (\bibinfo{date}{oct} \bibinfo{year}{2019}),
  \bibinfo{numpages}{27}~pages.
\newblock
\urldef\tempurl%
\url{https://doi.org/10.1145/3360585}
\showDOI{\tempurl}


\bibitem[Bavishi et~al\mbox{.}(2019)]%
        {bavishi2019phoenix}
\bibfield{author}{\bibinfo{person}{Rohan Bavishi}, \bibinfo{person}{Hiroaki
  Yoshida}, {and} \bibinfo{person}{Mukul~R. Prasad}.}
  \bibinfo{year}{2019}\natexlab{}.
\newblock \showarticletitle{Phoenix: Automated Data-Driven Synthesis of Repairs
  for Static Analysis Violations}. In \bibinfo{booktitle}{\emph{Proceedings of
  the 2019 27th ACM Joint Meeting on European Software Engineering Conference
  and Symposium on the Foundations of Software Engineering}} (Tallinn, Estonia)
  \emph{(\bibinfo{series}{ESEC/FSE 2019})}. \bibinfo{publisher}{Association for
  Computing Machinery}, \bibinfo{address}{New York, NY, USA},
  \bibinfo{pages}{613–624}.
\newblock
\showISBNx{9781450355728}
\urldef\tempurl%
\url{https://doi.org/10.1145/3338906.3338952}
\showDOI{\tempurl}


\bibitem[Chen et~al\mbox{.}(2021)]%
        {Codex}
\bibfield{author}{\bibinfo{person}{Mark Chen}, \bibinfo{person}{Jerry Tworek},
  \bibinfo{person}{Heewoo Jun}, \bibinfo{person}{Qiming Yuan},
  \bibinfo{person}{Henrique~Ponde de Oliveira~Pinto}, \bibinfo{person}{Jared
  Kaplan}, \bibinfo{person}{Harrison Edwards}, \bibinfo{person}{Yuri Burda},
  \bibinfo{person}{Nicholas Joseph}, \bibinfo{person}{Greg Brockman},
  \bibinfo{person}{Alex Ray}, \bibinfo{person}{Raul Puri},
  \bibinfo{person}{Gretchen Krueger}, \bibinfo{person}{Michael Petrov},
  \bibinfo{person}{Heidy Khlaaf}, \bibinfo{person}{Girish Sastry},
  \bibinfo{person}{Pamela Mishkin}, \bibinfo{person}{Brooke Chan},
  \bibinfo{person}{Scott Gray}, \bibinfo{person}{Nick Ryder},
  \bibinfo{person}{Mikhail Pavlov}, \bibinfo{person}{Alethea Power},
  \bibinfo{person}{Lukasz Kaiser}, \bibinfo{person}{Mohammad Bavarian},
  \bibinfo{person}{Clemens Winter}, \bibinfo{person}{Philippe Tillet},
  \bibinfo{person}{Felipe~Petroski Such}, \bibinfo{person}{Dave Cummings},
  \bibinfo{person}{Matthias Plappert}, \bibinfo{person}{Fotios Chantzis},
  \bibinfo{person}{Elizabeth Barnes}, \bibinfo{person}{Ariel Herbert{-}Voss},
  \bibinfo{person}{William~Hebgen Guss}, \bibinfo{person}{Alex Nichol},
  \bibinfo{person}{Alex Paino}, \bibinfo{person}{Nikolas Tezak},
  \bibinfo{person}{Jie Tang}, \bibinfo{person}{Igor Babuschkin},
  \bibinfo{person}{Suchir Balaji}, \bibinfo{person}{Shantanu Jain},
  \bibinfo{person}{William Saunders}, \bibinfo{person}{Christopher Hesse},
  \bibinfo{person}{Andrew~N. Carr}, \bibinfo{person}{Jan Leike},
  \bibinfo{person}{Joshua Achiam}, \bibinfo{person}{Vedant Misra},
  \bibinfo{person}{Evan Morikawa}, \bibinfo{person}{Alec Radford},
  \bibinfo{person}{Matthew Knight}, \bibinfo{person}{Miles Brundage},
  \bibinfo{person}{Mira Murati}, \bibinfo{person}{Katie Mayer},
  \bibinfo{person}{Peter Welinder}, \bibinfo{person}{Bob McGrew},
  \bibinfo{person}{Dario Amodei}, \bibinfo{person}{Sam McCandlish},
  \bibinfo{person}{Ilya Sutskever}, {and} \bibinfo{person}{Wojciech Zaremba}.}
  \bibinfo{year}{2021}\natexlab{}.
\newblock \showarticletitle{Evaluating Large Language Models Trained on Code}.
\newblock \bibinfo{journal}{\emph{CoRR}}  \bibinfo{volume}{abs/2107.03374}
  (\bibinfo{year}{2021}).
\newblock


\bibitem[Gao et~al\mbox{.}(2016)]%
        {bovinspector}
\bibfield{author}{\bibinfo{person}{F. Gao}, \bibinfo{person}{L. Wang}, {and}
  \bibinfo{person}{X. Li}.} \bibinfo{year}{2016}\natexlab{}.
\newblock \showarticletitle{BovInspector: Automatic inspection and repair of
  buffer overflow vulnerabilities}. In \bibinfo{booktitle}{\emph{2016 31st
  IEEE/ACM International Conference on Automated Software Engineering (ASE)}}.
  \bibinfo{publisher}{IEEE Computer Society}, \bibinfo{address}{Los Alamitos,
  CA, USA}, \bibinfo{pages}{786--791}.
\newblock
\urldef\tempurl%
\url{https://doi.ieeecomputersociety.org/}
\showURL{%
\tempurl}


\bibitem[Gazzola et~al\mbox{.}(2019)]%
        {survey1}
\bibfield{author}{\bibinfo{person}{L. Gazzola}, \bibinfo{person}{D. Micucci},
  {and} \bibinfo{person}{L. Mariani}.} \bibinfo{year}{2019}\natexlab{}.
\newblock \showarticletitle{Automatic Software Repair: A Survey}.
\newblock \bibinfo{journal}{\emph{IEEE Transactions on Software Engineering}}
  \bibinfo{volume}{45}, \bibinfo{number}{01} (\bibinfo{date}{jan}
  \bibinfo{year}{2019}), \bibinfo{pages}{34--67}.
\newblock
\showISSN{1939-3520}
\urldef\tempurl%
\url{https://doi.org/10.1109/TSE.2017.2755013}
\showDOI{\tempurl}


\bibitem[Huang et~al\mbox{.}(2019)]%
        {senx}
\bibfield{author}{\bibinfo{person}{Zhen Huang}, \bibinfo{person}{David Lie},
  \bibinfo{person}{Gang Tan}, {and} \bibinfo{person}{Trent Jaeger}.}
  \bibinfo{year}{2019}\natexlab{}.
\newblock \showarticletitle{Using Safety Properties to Generate Vulnerability
  Patches}. In \bibinfo{booktitle}{\emph{2019 IEEE Symposium on Security and
  Privacy (SP)}}. \bibinfo{pages}{539--554}.
\newblock
\urldef\tempurl%
\url{https://doi.org/10.1109/SP.2019.00071}
\showDOI{\tempurl}


\bibitem[Ke et~al\mbox{.}(2015)]%
        {SearchRepair:ASE2015}
\bibfield{author}{\bibinfo{person}{Yalin Ke}, \bibinfo{person}{Kathryn~T.
  Stolee}, \bibinfo{person}{Claire~Le Goues}, {and} \bibinfo{person}{Yuriy
  Brun}.} \bibinfo{year}{2015}\natexlab{}.
\newblock \showarticletitle{{Repairing Programs with Semantic Code Search
  (T)}}. In \bibinfo{booktitle}{\emph{Proceedings of the 2015 30th IEEE/ACM
  International Conference on Automated Software Engineering (ASE)}}
  \emph{(\bibinfo{series}{ASE '15})}. \bibinfo{publisher}{IEEE Computer
  Society}, \bibinfo{address}{Washington, DC, USA}, \bibinfo{pages}{295--306}.
\newblock


\bibitem[Kutsia et~al\mbox{.}(2011)]%
        {anti-unification}
\bibfield{author}{\bibinfo{person}{Temur Kutsia}, \bibinfo{person}{Jordi Levy},
  {and} \bibinfo{person}{Mateu Villaret}.} \bibinfo{year}{2011}\natexlab{}.
\newblock \showarticletitle{Anti-Unification for Unranked Terms and Hedges}. In
  \bibinfo{booktitle}{\emph{Proceedings of the 22nd International Conference on
  Rewriting Techniques and Applications, {RTA} 2011, May 30 - June 1, 2011,
  Novi Sad, Serbia}} \emph{(\bibinfo{series}{LIPIcs},
  Vol.~\bibinfo{volume}{10})}, \bibfield{editor}{\bibinfo{person}{Manfred
  Schmidt{-}Schau{\ss}}} (Ed.). \bibinfo{publisher}{Schloss Dagstuhl -
  Leibniz-Zentrum f{\"{u}}r Informatik}, \bibinfo{pages}{219--234}.
\newblock
\urldef\tempurl%
\url{https://doi.org/10.4230/LIPIcs.RTA.2011.219}
\showDOI{\tempurl}


\bibitem[Le~Goues et~al\mbox{.}(2012)]%
        {GenProg:ICSE2012}
\bibfield{author}{\bibinfo{person}{Claire Le~Goues}, \bibinfo{person}{Michael
  Dewey-Vogt}, \bibinfo{person}{Stephanie Forrest}, {and}
  \bibinfo{person}{Westley Weimer}.} \bibinfo{year}{2012}\natexlab{}.
\newblock \showarticletitle{{A Systematic Study of Automated Program Repair:
  Fixing 55 out of 105 Bugs for \$8 Each}}. In
  \bibinfo{booktitle}{\emph{Proceedings of the 34th International Conference on
  Software Engineering}} (Zurich, Switzerland) \emph{(\bibinfo{series}{ICSE
  '12})}. \bibinfo{publisher}{IEEE Press}, \bibinfo{address}{Piscataway, NJ,
  USA}, \bibinfo{pages}{3--13}.
\newblock


\bibitem[{Liu} et~al\mbox{.}(2018)]%
        {Liu:mining}
\bibfield{author}{\bibinfo{person}{K. {Liu}}, \bibinfo{person}{D. {Kim}},
  \bibinfo{person}{T.~F. {Bissyande}}, \bibinfo{person}{S. {Yoo}}, {and}
  \bibinfo{person}{Y. {Le Traon}}.} \bibinfo{year}{2018}\natexlab{}.
\newblock \showarticletitle{Mining Fix Patterns for FindBugs Violations}.
\newblock \bibinfo{journal}{\emph{IEEE Transactions on Software Engineering}}
  (\bibinfo{year}{2018}), \bibinfo{pages}{1--1}.
\newblock
\showISSN{0098-5589}
\urldef\tempurl%
\url{https://doi.org/10.1109/TSE.2018.2884955}
\showDOI{\tempurl}


\bibitem[Liu et~al\mbox{.}(2019)]%
        {liu2019avatar}
\bibfield{author}{\bibinfo{person}{Kui Liu}, \bibinfo{person}{Anil Koyuncu},
  \bibinfo{person}{Dongsun Kim}, {and} \bibinfo{person}{Tegawend{\'e}
  F.~Bissyand{\'e}}.} \bibinfo{year}{2019}\natexlab{}.
\newblock \showarticletitle{{AVATAR:} Fixing Semantic Bugs with Fix Patterns of
  Static Analysis Violations}. In \bibinfo{booktitle}{\emph{Proceedings of the
  26th IEEE International Conference on Software Analysis, Evolution, and
  Reengineering}}. IEEE, \bibinfo{pages}{456--467}.
\newblock


\bibitem[Long et~al\mbox{.}(2017)]%
        {genesis}
\bibfield{author}{\bibinfo{person}{Fan Long}, \bibinfo{person}{Peter Amidon},
  {and} \bibinfo{person}{Martin Rinard}.} \bibinfo{year}{2017}\natexlab{}.
\newblock \showarticletitle{Automatic Inference of Code Transforms for Patch
  Generation}. In \bibinfo{booktitle}{\emph{Proceedings of the 2017 11th Joint
  Meeting on Foundations of Software Engineering}} (Paderborn, Germany)
  \emph{(\bibinfo{series}{ESEC/FSE 2017})}. \bibinfo{publisher}{Association for
  Computing Machinery}, \bibinfo{address}{New York, NY, USA},
  \bibinfo{pages}{727–739}.
\newblock
\showISBNx{9781450351058}
\urldef\tempurl%
\url{https://doi.org/10.1145/3106237.3106253}
\showDOI{\tempurl}


\bibitem[Long and Rinard(2016)]%
        {Prophet}
\bibfield{author}{\bibinfo{person}{Fan Long} {and} \bibinfo{person}{Martin
  Rinard}.} \bibinfo{year}{2016}\natexlab{}.
\newblock \showarticletitle{{Automatic Patch Generation by Learning Correct
  Code}}. In \bibinfo{booktitle}{\emph{Proceedings of the 43rd Annual ACM
  SIGPLAN-SIGACT Symposium on Principles of Programming Languages}} (St.
  Petersburg, FL, USA) \emph{(\bibinfo{series}{POPL '16})}.
  \bibinfo{publisher}{ACM}, \bibinfo{address}{New York, NY, USA},
  \bibinfo{pages}{298--312}.
\newblock


\bibitem[Ma et~al\mbox{.}(2016)]%
        {cdrep}
\bibfield{author}{\bibinfo{person}{Siqi Ma}, \bibinfo{person}{David Lo},
  \bibinfo{person}{Teng Li}, {and} \bibinfo{person}{Robert~H. Deng}.}
  \bibinfo{year}{2016}\natexlab{}.
\newblock \showarticletitle{CDRep: Automatic Repair of Cryptographic Misuses in
  Android Applications}. In \bibinfo{booktitle}{\emph{Proceedings of the 11th
  ACM on Asia Conference on Computer and Communications Security}} (Xi'an,
  China) \emph{(\bibinfo{series}{ASIA CCS '16})}.
  \bibinfo{publisher}{Association for Computing Machinery},
  \bibinfo{address}{New York, NY, USA}, \bibinfo{pages}{711–722}.
\newblock
\showISBNx{9781450342339}
\urldef\tempurl%
\url{https://doi.org/10.1145/2897845.2897896}
\showDOI{\tempurl}


\bibitem[Ma et~al\mbox{.}(2017)]%
        {vurle}
\bibfield{author}{\bibinfo{person}{Siqi Ma}, \bibinfo{person}{Ferdian Thung},
  \bibinfo{person}{David Lo}, \bibinfo{person}{Cong Sun}, {and}
  \bibinfo{person}{Robert~H. Deng}.} \bibinfo{year}{2017}\natexlab{}.
\newblock \showarticletitle{VuRLE: Automatic Vulnerability Detection and Repair
  by Learning from Examples}. In \bibinfo{booktitle}{\emph{Computer Security --
  ESORICS 2017}}, \bibfield{editor}{\bibinfo{person}{Simon~N. Foley},
  \bibinfo{person}{Dieter Gollmann}, {and} \bibinfo{person}{Einar Snekkenes}}
  (Eds.). \bibinfo{publisher}{Springer International Publishing},
  \bibinfo{address}{Cham}, \bibinfo{pages}{229--246}.
\newblock
\showISBNx{978-3-319-66399-9}


\bibitem[Marcilio et~al\mbox{.}(2019)]%
        {sonarcube}
\bibfield{author}{\bibinfo{person}{D. Marcilio}, \bibinfo{person}{C.~A. Furia},
  \bibinfo{person}{R. Bonifacio}, {and} \bibinfo{person}{G. Pinto}.}
  \bibinfo{year}{2019}\natexlab{}.
\newblock \showarticletitle{Automatically Generating Fix Suggestions in
  Response to Static Code Analysis Warnings}. In \bibinfo{booktitle}{\emph{2019
  IEEE 19th International Working Conference on Source Code Analysis and
  Manipulation (SCAM)}}. \bibinfo{publisher}{IEEE Computer Society},
  \bibinfo{address}{Los Alamitos, CA, USA}, \bibinfo{pages}{34--44}.
\newblock
\urldef\tempurl%
\url{https://doi.org/10.1109/SCAM.2019.00013}
\showDOI{\tempurl}


\bibitem[Mesecan et~al\mbox{.}(2021)]%
        {hypergi}
\bibfield{author}{\bibinfo{person}{Ibrahim Mesecan}, \bibinfo{person}{Daniel
  Blackwell}, \bibinfo{person}{David Clark}, \bibinfo{person}{Myra Cohen},
  {and} \bibinfo{person}{Justyna Petke}.} \bibinfo{year}{2021}\natexlab{}.
\newblock \bibinfo{title}{HyperGI: Automated Detection and Repair of
  Information Flow Leakage}.
\newblock
\newblock


\bibitem[Monperrus(2020)]%
        {survey2}
\bibfield{author}{\bibinfo{person}{Martin Monperrus}.}
  \bibinfo{year}{2020}\natexlab{}.
\newblock \bibinfo{title}{{The Living Review on Automated Program Repair}}.
  (\bibinfo{date}{Dec.} \bibinfo{year}{2020}).
\newblock
\urldef\tempurl%
\url{https://hal.archives-ouvertes.fr/hal-01956501}
\showURL{%
\tempurl}
\newblock
\shownote{working paper or preprint}.


\bibitem[Pearce et~al\mbox{.}(2023)]%
        {LLMrepair}
\bibfield{author}{\bibinfo{person}{H. Pearce}, \bibinfo{person}{B. Tan},
  \bibinfo{person}{B. Ahmad}, \bibinfo{person}{R. Karri}, {and}
  \bibinfo{person}{B. Dolan-Gavitt}.} \bibinfo{year}{2023}\natexlab{}.
\newblock \showarticletitle{Examining Zero-Shot Vulnerability Repair with Large
  Language Models}. In \bibinfo{booktitle}{\emph{2023 2023 IEEE Symposium on
  Security and Privacy (SP) (SP)}}. \bibinfo{publisher}{IEEE Computer Society},
  \bibinfo{address}{Los Alamitos, CA, USA}, \bibinfo{pages}{1--18}.
\newblock
\urldef\tempurl%
\url{https://doi.org/10.1109/SP46215.2023.00001}
\showDOI{\tempurl}


\bibitem[Polikarpova et~al\mbox{.}(2020)]%
        {lifty}
\bibfield{author}{\bibinfo{person}{Nadia Polikarpova}, \bibinfo{person}{Deian
  Stefan}, \bibinfo{person}{Jean Yang}, \bibinfo{person}{Shachar Itzhaky},
  \bibinfo{person}{Travis Hance}, {and} \bibinfo{person}{Armando
  Solar-Lezama}.} \bibinfo{year}{2020}\natexlab{}.
\newblock \showarticletitle{Liquid Information Flow Control}.
\newblock \bibinfo{journal}{\emph{Proc. ACM Program. Lang.}}
  \bibinfo{volume}{4}, \bibinfo{number}{ICFP}, Article \bibinfo{articleno}{105}
  (\bibinfo{date}{aug} \bibinfo{year}{2020}), \bibinfo{numpages}{30}~pages.
\newblock
\urldef\tempurl%
\url{https://doi.org/10.1145/3408987}
\showDOI{\tempurl}


\bibitem[Rolim et~al\mbox{.}(2017)]%
        {rolim2017learning}
\bibfield{author}{\bibinfo{person}{Reudismam Rolim}, \bibinfo{person}{Gustavo
  Soares}, \bibinfo{person}{Loris D’Antoni}, \bibinfo{person}{Oleksandr
  Polozov}, \bibinfo{person}{Sumit Gulwani}, \bibinfo{person}{Rohit Gheyi},
  \bibinfo{person}{Ryo Suzuki}, {and} \bibinfo{person}{Björn Hartmann}.}
  \bibinfo{year}{2017}\natexlab{}.
\newblock \showarticletitle{Learning Syntactic Program Transformations from
  Examples}. In \bibinfo{booktitle}{\emph{ICSE 2017} (\bibinfo{edition}{icse
  2017} ed.)}.
\newblock


\bibitem[Sousa et~al\mbox{.}(2021)]%
        {revisar}
\bibfield{author}{\bibinfo{person}{Reudismam Sousa}, \bibinfo{person}{Gustavo
  Soares}, \bibinfo{person}{Rohit Gheyi}, \bibinfo{person}{Titus Barik}, {and}
  \bibinfo{person}{Loris D'Antoni}.} \bibinfo{year}{2021}\natexlab{}.
\newblock \showarticletitle{Learning Quick Fixes from Code Repositories}. In
  \bibinfo{booktitle}{\emph{Proceedings of the XXXV Brazilian Symposium on
  Software Engineering}} (Joinville, Brazil) \emph{(\bibinfo{series}{SBES
  '21})}. \bibinfo{publisher}{Association for Computing Machinery},
  \bibinfo{address}{New York, NY, USA}, \bibinfo{pages}{74–83}.
\newblock
\showISBNx{9781450390613}
\urldef\tempurl%
\url{https://doi.org/10.1145/3474624.3474650}
\showDOI{\tempurl}


\bibitem[van Tonder and Goues(2018)]%
        {FootPatch}
\bibfield{author}{\bibinfo{person}{Rijnard van Tonder} {and}
  \bibinfo{person}{Claire~Le Goues}.} \bibinfo{year}{2018}\natexlab{}.
\newblock \showarticletitle{Static Automated Program Repair for Heap
  Properties}. In \bibinfo{booktitle}{\emph{Proceedings of the 40th
  International Conference on Software Engineering}} (Gothenburg, Sweden)
  \emph{(\bibinfo{series}{ICSE '18})}. \bibinfo{publisher}{ACM},
  \bibinfo{address}{New York, NY, USA}, \bibinfo{pages}{151--162}.
\newblock
\showISBNx{978-1-4503-5638-1}
\urldef\tempurl%
\url{https://doi.org/10.1145/3180155.3180250}
\showDOI{\tempurl}


\bibitem[Vassena et~al\mbox{.}(2021)]%
        {blade}
\bibfield{author}{\bibinfo{person}{Marco Vassena}, \bibinfo{person}{Craig
  Disselkoen}, \bibinfo{person}{Klaus Gleissenthall}, \bibinfo{person}{Sunjay
  Cauligi}, \bibinfo{person}{Rami Kıcı}, \bibinfo{person}{Ranjit Jhala},
  \bibinfo{person}{Dean Tullsen}, {and} \bibinfo{person}{Deian Stefan}.}
  \bibinfo{year}{2021}\natexlab{}.
\newblock \showarticletitle{Automatically eliminating speculative leaks from
  cryptographic code with blade}.
\newblock \bibinfo{journal}{\emph{Proceedings of the ACM on Programming
  Languages}}  \bibinfo{volume}{5} (\bibinfo{date}{01} \bibinfo{year}{2021}),
  \bibinfo{pages}{1--30}.
\newblock
\urldef\tempurl%
\url{https://doi.org/10.1145/3434330}
\showDOI{\tempurl}


\bibitem[Wang et~al\mbox{.}(2021)]%
        {Wang2021CodeT5IU}
\bibfield{author}{\bibinfo{person}{Yue Wang}, \bibinfo{person}{Weishi Wang},
  \bibinfo{person}{Shafiq~R. Joty}, {and} \bibinfo{person}{Steven C.~H. Hoi}.}
  \bibinfo{year}{2021}\natexlab{}.
\newblock \showarticletitle{CodeT5: Identifier-aware Unified Pre-trained
  Encoder-Decoder Models for Code Understanding and Generation}.
\newblock \bibinfo{journal}{\emph{ArXiv}}  \bibinfo{volume}{abs/2109.00859}
  (\bibinfo{year}{2021}).
\newblock


\bibitem[Xia et~al\mbox{.}(2022)]%
        {Xia2022PracticalPR}
\bibfield{author}{\bibinfo{person}{Chun Xia}, \bibinfo{person}{Yuxiang Wei},
  {and} \bibinfo{person}{Lingming Zhang}.} \bibinfo{year}{2022}\natexlab{}.
\newblock \showarticletitle{Practical Program Repair in the Era of Large
  Pre-trained Language Models}.
\newblock \bibinfo{journal}{\emph{ArXiv}}  \bibinfo{volume}{abs/2210.14179}
  (\bibinfo{year}{2022}).
\newblock


\bibitem[Xin and Reiss(2017)]%
        {ssFix:ASE2017}
\bibfield{author}{\bibinfo{person}{Qi Xin} {and} \bibinfo{person}{Steven~P.
  Reiss}.} \bibinfo{year}{2017}\natexlab{}.
\newblock \showarticletitle{Leveraging Syntax-related Code for Automated
  Program Repair}. In \bibinfo{booktitle}{\emph{Proceedings of the 32Nd
  IEEE/ACM International Conference on Automated Software Engineering}}
  (Urbana-Champaign, IL, USA) \emph{(\bibinfo{series}{ASE 2017})}.
  \bibinfo{publisher}{IEEE Press}, \bibinfo{address}{Piscataway, NJ, USA},
  \bibinfo{pages}{660--670}.
\newblock


\bibitem[Xiong et~al\mbox{.}(2017)]%
        {ACS:ICSE2017}
\bibfield{author}{\bibinfo{person}{Yingfei Xiong}, \bibinfo{person}{Jie Wang},
  \bibinfo{person}{Runfa Yan}, \bibinfo{person}{Jiachen Zhang},
  \bibinfo{person}{Shi Han}, \bibinfo{person}{Gang Huang}, {and}
  \bibinfo{person}{Lu Zhang}.} \bibinfo{year}{2017}\natexlab{}.
\newblock \showarticletitle{Precise Condition Synthesis for Program Repair}. In
  \bibinfo{booktitle}{\emph{Proceedings of the 39th International Conference on
  Software Engineering}} (Buenos Aires, Argentina) \emph{(\bibinfo{series}{ICSE
  '17})}. \bibinfo{publisher}{IEEE Press}, \bibinfo{address}{Piscataway, NJ,
  USA}, \bibinfo{pages}{416--426}.
\newblock


\bibitem[Yasunaga and Liang(2020)]%
        {graphbased}
\bibfield{author}{\bibinfo{person}{Michihiro Yasunaga} {and}
  \bibinfo{person}{Percy Liang}.} \bibinfo{year}{2020}\natexlab{}.
\newblock \bibinfo{title}{Graph-based, Self-Supervised Program Repair from
  Diagnostic Feedback}.
\newblock
\newblock


\bibitem[Yasunaga and Liang(2021)]%
        {yasunaga2021break}
\bibfield{author}{\bibinfo{person}{Michihiro Yasunaga} {and}
  \bibinfo{person}{Percy Liang}.} \bibinfo{year}{2021}\natexlab{}.
\newblock \showarticletitle{Break-It-Fix-It: Unsupervised Learning for Program
  Repair}. In \bibinfo{booktitle}{\emph{International Conference on Machine
  Learning (ICML)}}.
\newblock


\bibitem[Zhang et~al\mbox{.}(2022a)]%
        {zhang2022overwatch}
\bibfield{author}{\bibinfo{person}{Yuhao Zhang}, \bibinfo{person}{Yasharth
  Bajpai}, \bibinfo{person}{Priyanshu Gupta}, \bibinfo{person}{Ameya Ketkar},
  \bibinfo{person}{Miltiadis Allamanis}, \bibinfo{person}{Titus Barik},
  \bibinfo{person}{Sumit Gulwani}, \bibinfo{person}{Arjun Radhakrishna},
  \bibinfo{person}{Mohammad Raza}, \bibinfo{person}{Gustavo Soares}, {and}
  \bibinfo{person}{Ashish Tiwari}.} \bibinfo{year}{2022}\natexlab{a}.
\newblock \showarticletitle{Overwatch: Learning Patterns in Code Edit
  Sequences}.
\newblock \bibinfo{journal}{\emph{CoRR}}  \bibinfo{volume}{abs/2207.12456}
  (\bibinfo{year}{2022}).
\newblock
\urldef\tempurl%
\url{https://doi.org/10.48550/arXiv.2207.12456}
\showDOI{\tempurl}
\showeprint[arXiv]{2207.12456}


\bibitem[Zhang et~al\mbox{.}(2022b)]%
        {exampleBasedJava}
\bibfield{author}{\bibinfo{person}{Ying Zhang}, \bibinfo{person}{Ya Xiao},
  \bibinfo{person}{Md~Mahir~Asef Kabir}, \bibinfo{person}{Danfeng~(Daphne)
  Yao}, {and} \bibinfo{person}{Na Meng}.} \bibinfo{year}{2022}\natexlab{b}.
\newblock \showarticletitle{Example-Based Vulnerability Detection and Repair in
  Java Code}. In \bibinfo{booktitle}{\emph{Proceedings of the 30th IEEE/ACM
  International Conference on Program Comprehension}} (Virtual Event)
  \emph{(\bibinfo{series}{ICPC '22})}. \bibinfo{publisher}{Association for
  Computing Machinery}, \bibinfo{address}{New York, NY, USA},
  \bibinfo{pages}{190–201}.
\newblock
\showISBNx{9781450392983}
\urldef\tempurl%
\url{https://doi.org/10.1145/3524610.3527895}
\showDOI{\tempurl}


\end{thebibliography}

\end{document}